\documentclass[12pt,english,php]{revtex4}
\usepackage[T1]{fontenc}
\usepackage[latin9]{inputenc}
\usepackage[letterpaper]{geometry}
\usepackage{color}
\usepackage{amsbsy}
\usepackage{amstext}
\usepackage{amssymb}
\usepackage{graphicx}
\usepackage{setspace}
\usepackage{esint}

\makeatletter
\@ifundefined{textcolor}{}
{%
 \definecolor{BLACK}{gray}{0}
 \definecolor{WHITE}{gray}{1}
 \definecolor{RED}{rgb}{1,0,0}
 \definecolor{GREEN}{rgb}{0,1,0}
 \definecolor{BLUE}{rgb}{0,0,1}
 \definecolor{CYAN}{cmyk}{1,0,0,0}
 \definecolor{MAGENTA}{cmyk}{0,1,0,0}
 \definecolor{YELLOW}{cmyk}{0,0,1,0}
}

\makeatother

\usepackage{babel}
\begin{document}

\title{Real frequency tearing layers with parallel dynamics and the effect
on locking and resistive wall modes }

\author{J.~M.~Finn$^{1}$, A.~J.~Cole$^{2}$, and D.~P.~Brennan$^{3}$}

\affiliation{$1$ Los Alamos National Laboratory, Los Alamos, NM 87545; Current
address: Tibbar Plasma Technologies, 274 DP Rd, Los Alamos, NM 87544 }

\affiliation{$2$ Dept.~of Applied Physics and Applied Mathematics, Columbia
University, New York, NY }

\affiliation{$3$- Dept.~of Astrophysical Sciences, Princeton University, Princeton,
NJ}
\begin{abstract}
\begin{singlespace}
Tearing modes with real frequencies in the plasma frame (i.e.~in
addition to the Doppler shift due to $E\times B$ rotation) are of
potential importance because of their effect on the locking process.
In particular, it has recently been shown \cite{FinnColeBrennan}
that the Maxwell torque on the plasma in the presence of an applied
error field is modified significantly for tearing modes having real
frequencies near marginal stability. In addition, it is known \cite{FinnGerwinModeCoupling}
that resistive wall tearing modes can be destabilized below their
no-wall limits by rotation, if the tearing modes have real frequencies
near marginal stability. In this paper we first derive the tearing
mode dispersion relation with pressure gradient, field line curvature
and parallel dynamics in the resistive-inertial (RI) regime, neglecting
the divergence of the $E\times B$ drift and perpendicular resistivity.
The results show that the usual Glasser effect, a toroidal effect
which involves real frequencies, occurs in this simplified model,
which ignores perpendicular resistivity and the divergence of the
$E\times B$ drift. We also find, using a similar simple model, the
surprising result that in the viscoresistive regime with pressure
gradient, favorable curvature due to toroidal effects, and parallel
dynamics, a similar Glasser-like effect is found. We show that in
both regimes the existence of tearing modes with complex frequencies
is related to nearby electrostatic resistive interchange modes with
complex frequencies. We discuss the effect on locking to an error
field and the significant lowering of the threshold for destabilization
of resistive wall tearing modes, which can be much more pronounced
than the weak effect observed for RI tearing modes without pressure-curvature
drive in Ref.~\cite{FinnGerwinModeCoupling}.\end{singlespace}

\end{abstract}
\maketitle

\section{Introduction\label{sec:Introduction}}

It is known that tearing modes can have real frequencies due to diamagnetism
\cite{Coppi-1,Coppi-2,Biskamp,FMA} or pressure gradient and favorable
curvature in the tearing layer, the Glasser effect \cite{GGJ1,GGJ2}.
It has recently been discovered that the process of locking to an
externally applied error field is affected strongly if the related
spontaneous tearing modes have real frequencies \cite{FinnColeBrennan}.
Specifically, for a nearly marginally stable mode with real frequency
$\omega_{r}$ in the plasma frame, $E\times B$ rotation leads to
a peak in the reconnected flux near at the value of rotation $v$
such that the mode has zero frequency in the laboratory frame $\omega_{r}+kv=0$.
That is, the flux is maximum when $v=-\omega_{r}/k$. Also, the Maxwell
torque applied at the tearing layer is zero for $v=-\omega_{r}/k$
rather than zero at $v=0$%
\footnote{The maximum of the reconnection flux and the zero of the torque curve
occur close to $v=-\omega_{r}/k$, but exactly at this point only
in the limit of growth rate $\gamma\rightarrow0-$ \cite{FinnColeBrennan}.%
}. For this reason, the plasma should lock to the phase velocity of
the tearing mode rather than to zero velocity \cite{FinnColeBrennan}.

It is perhaps less appreciated that the qualitative behavior of tearing
modes with real frequencies depends critically on parallel dynamics.
For example, the Glasser effect, with pressure gradient and favorable
field line curvature, is based on an intricate calculation involving
full resistive MHD including parallel dynamics \cite{GGJ1,GGJ2,FinnManheimer}.
In this paper we show a streamlined derivation of the Glasser effect
in the resistive-inertial (RI) tearing regime, in which ion inertia,
but not ion viscosity, is included in resistive MHD. This simplified
calculation based on reduced MHD includes parallel dynamics but neglects
the divergence of the $E\times B$ drift $\nabla\cdot\mathbf{v}_{\perp}$
and the perpendicular resistivity $\eta_{\perp}$. The purpose of
this derivation is threefold: (1) to show that these two last effects
are not necessary to obtain the qualitative results, i.e.~complex
roots and stabilization for positive constant-$\psi$ matching parameter
$\Delta'$; (2) to illustrate this simple approach for use in other
tearing regimes; and (3) to exploit the simplicity of this model to
elucidate the physics. We use this simplified method to investigate
the viscoresistive (VR) regime, which has perpendicular ion viscosity
but neglects perpendicular ion inertia in the resistive MHD layer.
These calculations include pressure gradient, favorable or unfavorable
curvature and parallel dynamics.

In Sec.~\ref{sec:Resistive-inertial-(RI)-regime} we show our derivation
of the RI regime in the presence of equilibrium pressure gradient
and favorable or unfavorable field line curvature in the tearing layer.
We show that the qualitative form for the tearing dispersion relation
in this regime is obtained by adding only parallel dynamics and ignoring
divergence of the $E\times B$ drift and the perpendicular resistivity
(classical particle transport). The equations are derived by an extended
reduced MHD formulation, in which specific effects are added one by
one to the most primitive form of reduced MHD, making the terms representing
the various physical effects evident. We have investigated the spontaneous
tearing response, showing nonmonotonic behavior of the inner layer
matching parameter $\Delta(Q)$ {[}c.f.~Eq.~(\ref{eq:RI-final-for-full-Delta(Q)}){]}.
This behavior leads to the well-known complex roots, which occur in
complex conjugate pairs. In addition to showing that $\nabla\cdot\tilde{\mathbf{v}}_{E}$
and $\eta_{\perp}$ are not required to obtain the qualitative behavior,
the results of this section show how the relevant effects, e.g.~pressure
gradient and curvature and parallel dynamics, are added easily and
transparently. In particular, we discuss the connection between the
complex roots and nearby stable electrostatic resistive interchange
modes with complex frequencies.

In Sec.~\ref{sec:Viscoresistive-(VR)-regime}, we formulate the VR
regime with pressure gradient, favorable curvature and parallel dynamics.
We again neglect the divergence of the $E\times B$ drift and perpendicular
resistivity, and use the methods outlined in Sec.~\ref{sec:Resistive-inertial-(RI)-regime}.
The results show, surprisingly, that there is a large range of parameters
for which $\Delta(Q)$ is again nonmonotonic, and this behavior again
leads to complex conjugate roots and stabilization. We again explore
the relationship between the electrostatic resistive interchange modes
and (electromagnetic) tearing modes with complex frequencies. 

In Sec.~\ref{sub:Resonant-field-amplification} we discuss error
field penetration and locking for tearing layers whose spontaneous
modes have non-zero real frequencies, e.g.~the RI and VR models with
Glasser effect. We also discuss, in Sec.~\ref{sub:Resistive-wall-tearing},
resistive wall tearing modes with such tearing layers. We show that
for the ideal wall tearing mode near marginal stability with real
frequency due to the Glasser effect, the resistive wall mode is destabilized
below the no wall tearing threshold much more noticeably than in the
RI regime without pressure-curvature drive \cite{FinnGerwinModeCoupling}.

In Sec.~\ref{sec:Conclusions} we summarize our results. On the basis
of these results, we suggest that tearing modes typically have complex
frequencies and their effects, namely finite velocity locking and
destabilization of resistive wall modes for small $E\times B$ rotation.
Lastly, we discuss other tearing regimes, of more possible relevance
to high temperature plasmas, with complex frequencies.

\section{Resistive-inertial (RI) regime with pressure gradient and curvature\label{sec:Resistive-inertial-(RI)-regime}}

In this section we review the resistive-inertial (RI) tearing regime.
We describe the equations in terms of ``extended reduced resistive
MHD'', i.e.~we include effects beyond the most basic reduced resistive
MHD model one at a time. In the first subsection we include the pressure-curvature
term in the vorticity equation from the term proportional to $\mathbf{B}\times\boldsymbol{\kappa}\cdot\nabla p$,
where $\boldsymbol{\boldsymbol{\kappa}}=\hat{\mathbf{b}}\cdot\nabla\hat{\mathbf{b}}$
is the field line curvature, using the advected pressure model ($\Gamma p_{0}\rightarrow0$
in the adiabatic law.) The reduced resistive MHD model for tokamaks
applies for large aspect ratio $R/r\gg1$ so that, with $q(r)=rB_{z}/RB_{\theta}\gtrsim1$,
we can conclude that $B_{z}$ is large and constant, $B_{z}=B_{0}$.
Toroidal geometry, specifically toroidal curvature of the field lines,
is included to the degree that the average field line curvature in
the layer, favorable where $q>1$, is included. We proceed to add
parallel dynamics in the next subsection. The purpose of this section
is to rederive known results with a simple and transparent model,
so that it can be used to obtain new results in other regimes.

\subsection{RI regime without parallel dynamics\label{sub:RI-regime-without-parallel}}

The vorticity, parallel Ohm's law, and the pressure equation, with
$\tilde{\mathbf{B}}=\nabla\tilde{\psi}\times\hat{\mathbf{z}}$, $\tilde{\mathbf{v}}_{\perp}=\nabla\tilde{\phi}\times\hat{\mathbf{z}}$
in the RI regime (including ion inertia but with zero ion viscosity
$\mu=0$) take the form
\begin{equation}
\rho\gamma\nabla_{\perp}^{2}\tilde{\phi}=iF\nabla_{\perp}^{2}\tilde{\psi}+\frac{2imB_{\theta}^{2}}{B_{0}^{2}r^{2}}\tilde{p},\label{eq:vorticity}
\end{equation}

\begin{equation}
\gamma\tilde{\psi}=iF\tilde{\phi}+\eta\nabla_{\perp}^{2}\tilde{\psi},\,\,\,\,\,\,\,\text{(a)}\,\qquad\gamma\tilde{p}=-\frac{imp'}{r}\tilde{\phi},\,\,\,\,\,\,\,\text{(b)}\label{eq:Ohm-adiabatic-bare}
\end{equation}
where $F(r)=\mathbf{k}\cdot\mathbf{B}=mB_{\theta}/r+kB_{0}$, $B_{z}=B_{0}$
is constant and the compression term proportional to $\Gamma p_{0}$
is not included. Close to the layer at $r=r_{t}$ we have $F(r)=\alpha x$
with $x=r-r_{t}$ and $\alpha=F'(r_{t})=(B_{\theta}/r)(m-nq(r))'=-(nB_{\theta}/r)q'(r_{t})$.
Again, we write $q=rB_{0}/RB_{\theta}$, and in cylindrical geometry
the mode behaves as $\tilde{\mathbf{B}}=\tilde{\mathbf{B}}(r)e^{im\theta+ikz}$,
with $k=-n/R$. In the close-in outer region, i.e.~neglecting inertia
and resistivity but still near the layer, we find
\[
x(x\tilde{\phi})''-\frac{2m^{2}p'(r_{t})}{B_{0}^{2}r_{t}n^{2}q'(r_{t})^{2}}\tilde{\phi}=0,
\]
or
\begin{equation}
(x^{2}\tilde{\phi}')'+D_{s}\tilde{\phi}=0,\label{eq:Outer-asymptotic}
\end{equation}
where 
\begin{equation}
D_{s}=-\frac{2m^{2}p'(r_{t})}{B_{0}^{2}r_{t}n^{2}q'(r_{t})^{2}}=-\frac{2r_{t}p'(r_{t})}{B_{\theta}^{2}R^{2}q'(r_{t})^{2}}\label{eq:Suydam}
\end{equation}
is the usual Suydam parameter and we have used $q(r_{t})=m/n$. In
toroidal geometry this is replaced by the Mercier parameter \cite{GGJ1,GGJ2}
\begin{equation}
D=\left(1-q(r_{t})^{2}\right)D_{s}.\label{eq:Mercier}
\end{equation}
For $p'(r_{t})<0$ and $q(r_{t})>1$ we have favorable curvature,
$D<0$. For normal profiles, we have $|D|\sim|D_{s}|\sim\beta\sim2p/B_{0}^{2}\ll1$.
We will be more specific about the ordering of $\beta$ below.\textcolor{red}{{}
}From Eq.~(\ref{eq:Outer-asymptotic}) with $D_{s}\rightarrow D$
we have $\tilde{\phi}\propto x^{p}$ with $p=\left(-1\pm\sqrt{1-4D}\right)/2$.
For $D\ll1$ we have $p=-1,0$ and we can use the constant-$\psi$
approximation in the usual (zero pressure) way.

In the tearing layer we have
\begin{equation}
\rho\gamma\tilde{\phi}''=i\alpha x\tilde{\psi}''-E\tilde{\phi},\qquad\gamma\tilde{\psi}=i\alpha x\tilde{\phi}+\eta\tilde{\psi}'',\label{eq:RI-2-equations}
\end{equation}
where $E=-2m^{2}B_{\theta}^{2}(r_{t})p'(r_{t})/\gamma B_{0}^{2}r_{t}^{3}=\alpha^{2}D_{s}/\gamma=E_{0}/\gamma$
and again toroidal geometry leads to $E_{0}\rightarrow\left(1-q(r_{t})^{2}\right)E_{0}$.
Substituting, we find
\begin{equation}
\rho\gamma\tilde{\phi}''-\frac{\alpha^{2}x^{2}}{\eta}\tilde{\phi}+E\tilde{\phi}=\frac{i\alpha\gamma}{\eta}x\tilde{\psi}_{0},\label{eq:RawMomentumSubstituted}
\end{equation}
where $\tilde{\psi}_{0}$ is the lowest order, constant part of $\tilde{\psi}$.

We let $x=\delta\xi$, defining $\delta$ by the first two terms,
$\delta^{4}\alpha{}^{2}=\rho\gamma\eta$. With $\tilde{\phi}=-i\alpha\delta^{3}\tilde{\psi}_{0}W/\rho\eta=-i\gamma\tilde{\psi}_{0}W(\xi)/\alpha\delta$
we obtain
\begin{equation}
\frac{d^{2}W}{d\xi^{2}}-\xi^{2}W+GW=-\xi,\label{eq:2nd-Order-with-G}
\end{equation}
 where 
\begin{equation}
G=\frac{E\delta^{2}}{\rho\gamma},\label{eq:G-def-RI}
\end{equation}
and in toroidal geometry $G$ can contain the factor $1-q(r_{t})^{2}$.
The scalings $E=E_{0}/\gamma$, $\delta\sim\gamma^{1/4}$ lead to
$G\sim\gamma^{-3/2}$.

We do the usual constant-$\psi$ matching to find
\[
\Delta'=\Delta(\gamma)=\frac{[\tilde{\psi}']}{\tilde{\psi}_{0}}=\frac{1}{\eta\tilde{\psi}_{0}}\int\left(\gamma\tilde{\psi}_{0}-i\alpha x\tilde{\phi}\right)dx\,\,\,\,\,\text{or}
\]
\begin{equation}
\Delta'=\Delta(\gamma)=\frac{\delta\gamma}{\eta}\int_{-\infty}^{\infty}\left(1-\xi W\right)d\xi,\label{eq:Delta'Matching}
\end{equation}
where $\Delta'$ is obtained from the outer region. Then $\delta=\left(\rho\gamma\eta/\alpha^{2}\right)^{1/4}$
gives the RI constant-$\psi$ dispersion relation

\begin{equation}
\Delta'=\left(\frac{\gamma^{5}\rho}{\alpha^{2}\eta^{3}}\right)^{1/4}\Delta_{s}=(\gamma\tau_{ri})^{5/4}\Delta_{s},\,\,\,\,\,\,\,\Delta_{s}\equiv\int_{-\infty}^{\infty}\left(1-\xi W\right)d\xi.\label{eq:Delta_SDefinition}
\end{equation}

The constant-$\psi$ ordering parameter is $\Delta'\delta=\epsilon$
and we have $\delta\sim\epsilon$, $\gamma\sim\epsilon^{3/2}$. Notice
that in Eq.~(\ref{eq:2nd-Order-with-G}) the behavior $W\rightarrow1/\xi$
as $\xi\rightarrow\infty$ is not affected by the presence of $G$,
so that the integral for $\Delta_{s}$ still converges. For zero pressure
we have the standard result $\Delta_{s}=2.12$. Otherwise, $W$ is
influenced by $G$ and therefore $\Delta_{s}$ is changed. We have
defined $\gamma_{ri}=1/\tau_{ri}$ by $\gamma_{ri}^{5}\rho/\alpha^{2}\eta^{3}=1$
or $\gamma_{ri}=\left(\eta^{3}\alpha^{2}/\rho\right)^{1/5}$ and $\gamma=\gamma_{ri}Q$
or $\gamma\tau_{ri}=Q$, with $Q\sim1$. We have $\delta^{2}/\rho\gamma^{2}=\left(\eta/\alpha^{2}\rho(\gamma_{ri}Q)^{3}\right)^{1/2}$.
With the above relations we obtain $\delta^{2}/\rho\gamma_{ri}^{2}=\left(1/\eta^{2}\rho\alpha^{8}\right)^{1/5}$
and
\begin{equation}
G=-\frac{2m^{2}B_{\theta}^{2}(r_{t})p'(r_{t})}{B_{0}^{2}r_{t}^{3}}\frac{\delta^{2}}{\rho\gamma_{ri}^{2}}\frac{1}{Q^{3/2}}=\frac{G_{0}}{Q^{3/2}},\label{eq:G-->G0-RI}
\end{equation}
consistent with the scaling above. With these modifications, Eq.~(\ref{eq:Delta_SDefinition})
becomes
\begin{equation}
\Delta'=Q^{5/4}\Delta_{s}(Q)\label{eq:Delta'=00003Dq54Deltas}
\end{equation}
and the effect of pressure is contained in $\Delta_{s}(Q)$. The ordering
$\beta\sim\epsilon$, with $\delta\sim\epsilon\sim\eta^{2/5}$ and
$\gamma\sim\epsilon^{3/2}$ shows $G_{0}=O(1)$. (This assumes $B_{\theta}/B_{0}\sim r_{t}/R$,
which is small but of order unity with respect to $\epsilon$.)

As expected, the growth rate of the modes decreases for favorable
curvature, $G_{0}<0$. That is, as $G_{0}<0$ decreases $\Delta_{s}(Q)$
increases, so the solution $Q$ to Eq.~(\ref{eq:Delta'=00003Dq54Deltas}),
decreases. (Not shown.) Similarly, for $G_{0}>0$, increasing $G_{0}$
causes an increase in $Q$. 

In addition to the tearing modes which connect to the outer region
($\tilde{\psi}_{0}\neq0$; electromagnetic tearing modes), there are
localized electrostatic resistive interchange modes. These are solutions
to the homogeneous form of Eq.~(\ref{eq:2nd-Order-with-G}). For
$G_{0}>0$ (unfavorable curvature) these are Hermite functions $W(\xi)=H_{n}(\xi)e^{-\xi^{2}/2}$
with real eigenvalues 
\begin{equation}
Q_{n}=\left(\frac{G_{0}}{2n+1}\right)^{2/3},\label{eq:Hermite}
\end{equation}
where we have used Eq.~(\ref{eq:G-->G0-RI}). These electrostatic
interchanges have $\gamma_{es}=\gamma_{ri}Q_{n}$. This behavior $Q\sim|p'|^{2/3}$
is well known, and these eigenvalues have an accumulation point at
$Q=0$. Accordingly, the function $\Delta_{s}(Q)$ has poles on the
positive real axis at these eigenvalues for $G_{0}>0$, with an accumulation
point of poles $Q_{n}\rightarrow0+$ at $Q=0$. The poles in $\Delta_{s}(Q)$
correspond only to eigenvalues with odd $n$, for which $W$ is odd
in $\xi$. Similar behavior occurs in the work presented in Secs.~\ref{sub:RI-regime-with-parallel}
and \ref{sec:Viscoresistive-(VR)-regime}.) For $G_{0}<0$, the poles
are in the complex plane $Q\propto e^{\pm2\pi i/3}$ with $\text{Re}(Q)<0$,
corresponding to stable modes. Therefore, no clear evidence of these
poles shows up in plots of $\Delta_{s}(Q)$ or $\Delta(Q)$ on the
real $Q$ axis. We will return to the issue of these poles in Secs.~\ref{sub:RI-regime-with-parallel}
and \ref{sec:Viscoresistive-(VR)-regime}.

\subsection{RI regime with parallel dynamics\label{sub:RI-regime-with-parallel}}

To include parallel dynamics, we take the model above, add parallel
compression to the adiabatic law and include a parallel momentum equation,

\begin{equation}
\gamma\tilde{p}=-\frac{imp'}{r}\tilde{\phi}-ik_{||}\Gamma p\tilde{v}_{||},\label{eq:AdiabatWithParallel}
\end{equation}
\begin{equation}
\rho\gamma\tilde{v}_{||}=-ik_{||}\tilde{p}-\frac{imp'}{rB_{0}}\tilde{\psi}_{0},\label{eq:vParallelEq}
\end{equation}
where $k_{||}=F(r)/B_{0}$ and $\Gamma$ is the adiabatic index. Note
that we have included parallel inertia but not parallel viscosity.
In tearing modes driven by parallel velocity shear, the results using
either sound wave dynamics (parallel inertia and pressure gradient)
or parallel viscosity were found to be quite similar \cite{Finn-vparallel-prime}.
The $E\times B$ drift $\tilde{\mathbf{v}}_{\perp}=\nabla\tilde{\phi}\times\hat{\mathbf{z}}\rightarrow-\nabla\tilde{\Phi}\times\hat{\mathbf{z}}/B$,
where $\tilde{\Phi}\approx-B\tilde{\phi}$ is the electrostatic potential,
is incompressible in lowest order reduced MHD, where $B=|\mathbf{{B}|}$
equals $B_{0}$, which is constant. Extending reduced MHD to include
$\nabla B\neq0$ ($\nabla\cdot\tilde{\mathbf{v}}_{\perp}\approx-2\boldsymbol{\kappa}\cdot\nabla\tilde{\phi}\times\hat{\mathbf{z}}=-2\tilde{\mathbf{v}}_{\perp}\cdot\boldsymbol{\kappa}$)
and the perpendicular resistivity $\eta_{\perp}$ (particle transport)
gives all the terms in general compressible resistive MHD \cite{GGJ1,GGJ2,FinnManheimer}.
See Appendix A. We proceed without these effects and show that the
known qualitative behavior is present without the effects of $\nabla\cdot\tilde{\mathbf{v}}_{\perp}$
and $\eta_{\perp}$. We find 
\begin{equation}
\left(\begin{array}{cc}
\rho\gamma & ik_{||}\\
ik_{||}\Gamma p & \gamma
\end{array}\right)\left(\begin{array}{c}
\tilde{v}_{||}\\
\tilde{p}
\end{array}\right)=-\frac{imp'}{r}\left(\begin{array}{c}
\tilde{\psi}_{0}/B_{0}\\
\tilde{\phi}
\end{array}\right)\label{eq:Little-2X2-for-parallel}
\end{equation}
and thus
\begin{equation}
\tilde{p}=-\frac{imp'}{r}\frac{1}{\gamma^{2}+k_{||}^{2}c_{s}^{2}}\left(\gamma\tilde{\phi}-\frac{ik_{||}c_{s}^{2}}{B_{0}}\tilde{\psi}_{0}\right),\label{eq:Press-parallel-dyn}
\end{equation}
where $c_{s}^{2}=\Gamma p/\rho$ is the square of the sound speed.
Referring to Eq.~(\ref{eq:Ohm-adiabatic-bare}b), the first term
here has $\tilde{\phi}/\gamma\rightarrow\gamma\tilde{\phi}/\left(\gamma^{2}+k_{||}^{2}c_{s}^{2}\right)$,
and the sound wave propagator $\propto1/\left(\gamma^{2}+k_{||}^{2}c_{s}^{2}\right)$
represents the stabilizing influence of the pressure perturbation
leaking away along the field lines by sound wave propagation. The
second term in Eq.~(\ref{eq:Press-parallel-dyn}) represents the
tilting of the field lines by $\tilde{\psi}_{0}$ into the equilibrium
pressure gradient. This effect gives an additional pressure perturbation
of the same sign as the part proportional to $\tilde{\phi}$, e.g.~destabilizing
if $G_{0}>0$, but it is also mitigated by the sound wave propagator.
These parallel dynamics terms are important unless $\gamma^{2}\gg k_{||}'^{2}\delta^{2}c_{s}^{2}$,
when the advected pressure model is regained. Also, even if $\gamma^{2}\ll k_{||}'^{2}\delta^{2}c_{s}^{2}$,
the advected pressure model holds in the ideal MHD outer region, where
$\tilde{\psi}_{0}=ik_{||}B_{0}\tilde{\phi}/\gamma$.

Substituting (\ref{eq:Press-parallel-dyn}) for (\ref{eq:Ohm-adiabatic-bare}b)
and normalizing as before, we find that we must replace 
\begin{equation}
GW\longrightarrow\frac{GQ^{2}}{Q^{2}+b^{2}\xi^{2}}W+\frac{Gb^{2}}{Q^{2}+b^{2}\xi^{2}}\xi\label{eq:Replacement}
\end{equation}
in Eq.~(\ref{eq:2nd-Order-with-G}), where 
\begin{equation}
b^{2}=\alpha^{2}\delta^{2}c_{s}^{2}/\gamma_{ri}^{2}B_{0}^{2}=b_{0}^{2}Q^{1/2}\label{eq:RI-b-in-terms-of-b0}
\end{equation}
 is dimensionless and $O(\beta)$. The last relation defining $b_{0}$
is valid because $\delta\sim\gamma^{1/4}$ in the RI regime. We find
\begin{equation}
\frac{d^{2}W}{d\xi^{2}}-\xi^{2}W+\frac{Q^{2}}{Q^{2}+b^{2}\xi^{2}}GW=-\left(1+\frac{Gb^{2}}{Q^{2}+b^{2}\xi^{2}}\right)\xi.\label{eq:2ndOrderG-parallel}
\end{equation}
As discussed above, the sound wave reduction $Q^{2}/(Q^{2}+b^{2}\xi^{2})$
in the third term in (\ref{eq:2ndOrderG-parallel}) weakens the effect
of $G$ (e.g.~weakens the destabilizing effect for $G>0$) and the
other source of perturbed pressure, the term proportional to $G$
in the last term, is destabilizing for $G>0$, although it is also
weakened by sound wave propagation. According to Appendix A, the inclusion
of the divergence of the $E\times B$ drift effectively leads to a
small decrease in $G$ in the term proportional to $W$ with no change
in the term proportional to $G$ on the right. The inclusion of the
perpendicular resistivity increases the order and hence the complexity
of the equations, as we will discuss.

The relation above, $b^{2}=b_{0}^{2}Q^{1/2}$ and the equality $G=G_{0}/Q^{3/2}$
lead to a symmetry that leaves this equation and therefore $\Delta_{s}(Q)$
invariant: $G_{0}\rightarrow\lambda G_{0},\,\, b_{0}\rightarrow\mu b_{0},\,\, Q\rightarrow\nu Q$,
with $\lambda=\nu^{3/2},\,\,\mu^{2}=\nu^{3/2}$. The two group invariant
quantities are $G_{0}/b_{0}^{2}$ and $b_{0}^{2}/Q^{3/2}$, or equivalently
$G_{0}/b_{0}^{2}$ and $G_{0}/Q^{3/2}$; the latter is the quantity
$G$. The quantity $b_{0}^{2}/Q^{3/2}$ is the ratio of the second
to the first term in the propagator $Q^{2}+b^{2}\xi^{2}=Q^{2}+b_{0}^{2}Q^{1/2}\xi^{2}$
for $\xi^{2}\sim1$ and therefore measures the stabilization due to
sound wave propagation in the layer, i.e.~the reduction $Q^{2}/(Q^{2}+b^{2}\xi^{2})=\gamma^{2}/(\gamma^{2}+k_{||}^{2}c_{s}^{2})$.
The quantity $G_{0}/Q^{3/2}$ measures the inverse of the growth rate
relative to the growth rate of the electrostatic modes $\gamma_{es}$
from Eq.~(\ref{eq:Hermite}) $Q_{n}\sim G_{0}^{2/3}$. The quantity
$Gb^{2}/Q^{2}=\left(G_{0}/Q^{3/2}\right)\left(b_{0}^{2}/Q^{3/2}\right)$
measures the magnitude of the last term in Eq.~(\ref{eq:2ndOrderG-parallel})
without the sound wave reduction. Finally, the quantity $G_{0}/b_{0}^{2}$
equals $\left[2q(r_{t}^{2})/r_{t}^{2}q'(r_{t})^{2}\right]\left[-r_{t}p'(r_{t})/\Gamma p(r_{t})\right]$,
and typically both terms are of order unity.

\begin{figure}
\noindent\includegraphics[scale=0.5]{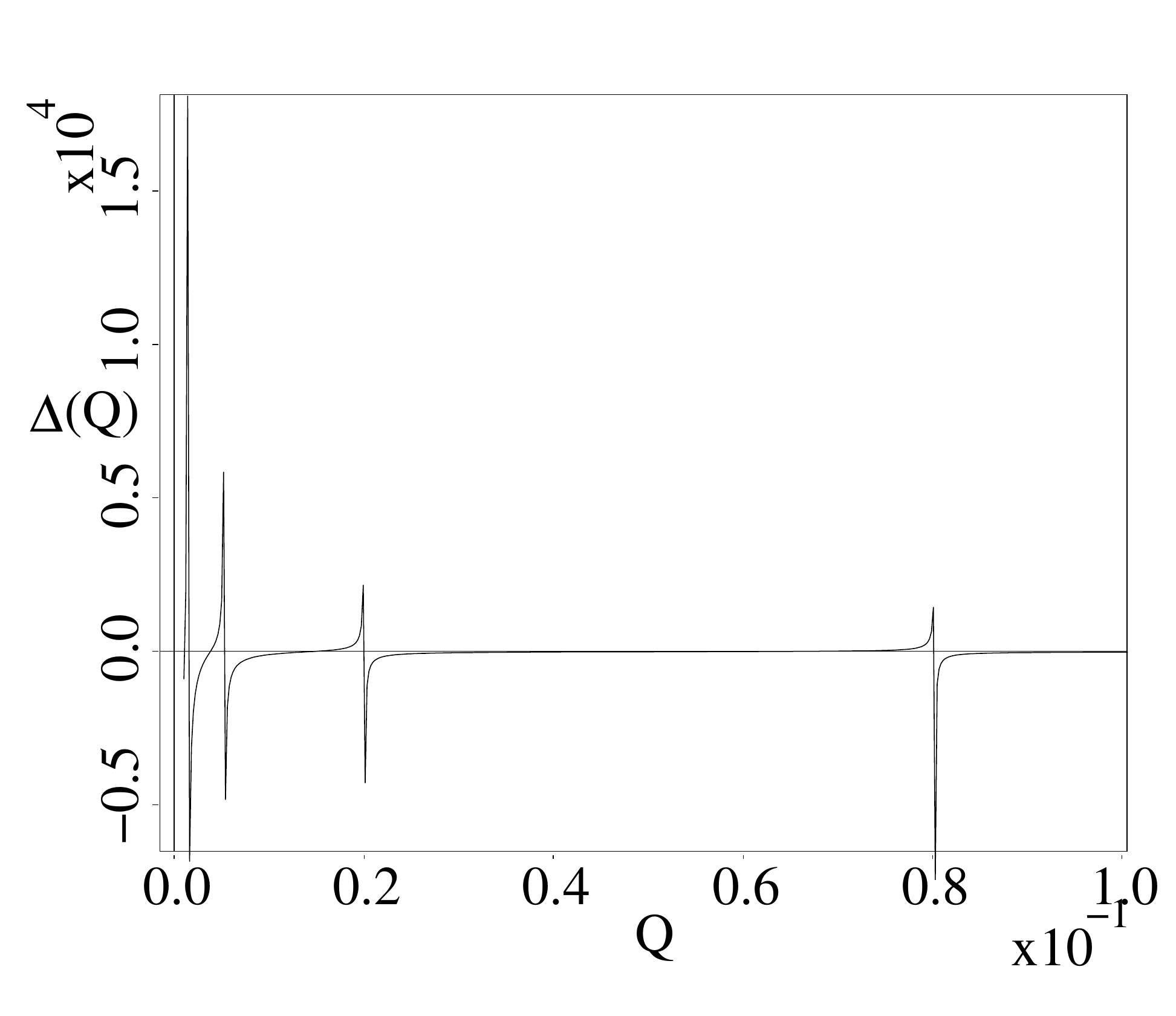} \raggedright{}\caption{
Plot of $\Delta_{s}(Q)$ in the RI regime with unfavorable curvature,
with small sound speed, $G_{0}=0.1,\, b_{0}=0.1$ ($g=G_{0}/b_{0}^{2}=10$),
showing four poles related to electrostatic resistive interchange
modes. \label{fig:Delta_S(Q)-curve-RI-4-poles}}
\end{figure}

\begin{figure}
\noindent\includegraphics[scale=0.5]{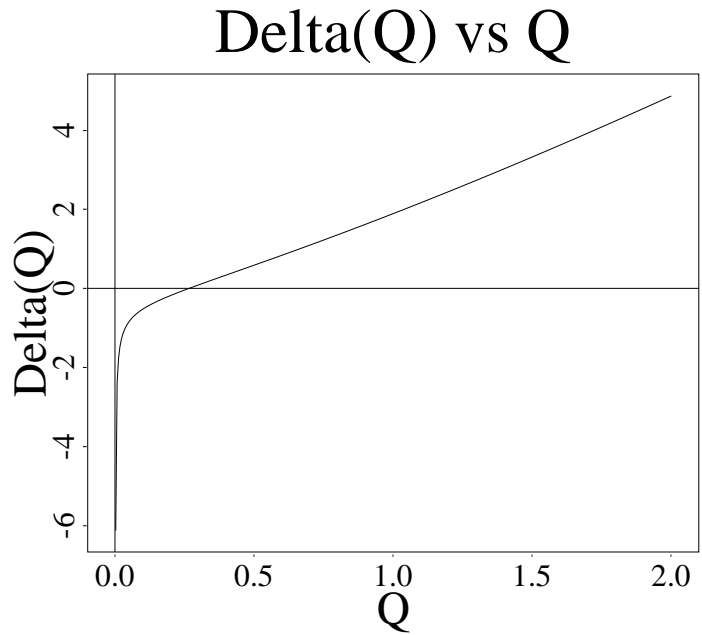}\caption{
Plot of $\Delta(Q)$ in the RI regime with unfavorable curvature and
fairly large sound speed (small $g=G_{0}/b_{0}^{2}$.) \label{fig:Delta(Q)-RI-FM}}
\end{figure}

We have found numerical solutions to Eq.~(\ref{eq:2ndOrderG-parallel})
and numerically evaluated the integral for $\Delta_{s}(Q)$ in Eq.~(\ref{eq:Delta_SDefinition}).
For $G_{0}$ positive (unfavorable curvature) and sufficiently large,
the $\Delta_{s}(Q)$ curve on the real $Q$ axis has poles, as shown
in Fig.~\ref{fig:Delta_S(Q)-curve-RI-4-poles}, corresponding to
localized unstable electrostatic resistive interchanges, at $Q=Q_{n}$,
the compressional analogs of the electrostatic modes discussed in
Sec.~\ref{sub:RI-regime-without-parallel}. 
All these modes are stabilized ($Q_{n}\rightarrow0$) for sufficiently
large sound speed ($G_{0}/b_{0}^{2}\propto p'/\Gamma p$ small enough.)
As we increase the sound speed (increase $b_{0}$ or decrease $g\equiv G_{0}/b_{0}^{2}$),
the last pole eventually goes through $Q=0$ into the complex plane
and we still observe $\Delta_{s}(Q)\rightarrow-\infty$ as $Q\rightarrow0$.
See Fig.~\ref{fig:Delta(Q)-RI-FM}. The singular behavior in this
figure at $Q=0$ appears to be a remnant of the electrostatic modes.
Since $W(\xi)$ is unchanged, and therefore $\Delta_{s}$ is unchanged,
if $G_{0},\, b_{0}^{2}$ and $Q$ are changed with $G=G_{0}/Q^{3/2}$
and $g=G_{0}/b_{0}^{2}$ fixed, we conclude that $\Delta_{s}$ depends
only on these two quantities, $\Delta_{s}=\Delta_{s}(G,g)$. For sufficiently
small $g$ we observe numerically $\Delta_{s}=C_{1}-C_{2}/Q^{3/2}$
for $G$ fixed, with $C_{1},\, C_{2}>0$ and $C_{2}\propto G_{0}$.
We thus can write 
\begin{equation}
\Delta_{s}(Q)=\Delta_{s}(Q=\infty)-\frac{G_{0}}{Q^{3/2}}K\left(\frac{G_{0}}{b_{0}^{2}}\right)\label{eq:RI-Delta_s-prelim}
\end{equation}
for some function $K(g)$. Expressing $K$ as a Taylor series in $g$,
we find a fit to the numerical data for large sound speed, small $g=G_{0}/b_{0}^{2}$,
\begin{equation}
\Delta_{s}(Q)=2.12-\frac{2.77G_{0}}{Q^{3/2}}\left(1+0.65\frac{G_{0}}{b_{0}^{2}}\right)\label{eq:RI-final-result}
\end{equation}
or
\begin{equation}
\Delta(Q)=2.12Q^{5/4}-\frac{2.77G_{0}}{Q^{1/4}}\left(1+0.65\frac{G_{0}}{b_{0}^{2}}\right).\label{eq:RI-final-for-full-Delta(Q)}
\end{equation}
This fit is excellent for $g\ll1$ and quite good for $g<0.2$. See
Fig.~\ref{fig:Delta(Q)-RI-FM}.

It is not difficult to show that $g=G_{0}/b_{0}^{2}$ is equal to
$2D_{s}/\Gamma\beta$, with $\beta=2p(r_{t})/B_{0}^{2}$. The variable
$g$ is the quantity that enters in Refs.~\cite{CGJ,FinnManheimer},
and we have $2D_{s}/\Gamma\beta=-2r_{t}p'(r_{t})/(\Gamma p(r_{t})s^{2})$,
where $s^{2}=\left(r_{t}q'(r_{t})/q(r_{t})\right)^{2}$ is the dimensionless
shear parameter, which is of order unity. It is reasonable to assume
$g\lesssim1$, and use the expansion in Eq.~(\ref{eq:RI-final-for-full-Delta(Q)}),
ignoring the perpendicular resistivity term. See Appendix A. Equation
(\ref{eq:RI-final-for-full-Delta(Q)}) shows the form $\Delta(Q)=C_{1}Q^{5/4}-C_{2}Q^{-1/4}$,
with $C_{1}>0,\, C_{2}>0$, as in Refs.~\cite{GGJ1,GGJ2,FinnManheimer}.
From this form of $\Delta(Q)$ (see Fig.~\ref{fig:Delta(Q)-RI-FM})
it is seen that the unfavorable curvature ($G_{0}>0$) results of
Ref.~\cite{FinnManheimer} are recovered, in particular $Q\sim(G_{0}/|\Delta'|)^{4}$
as $\Delta'\rightarrow-\infty$. It is clear that this behavior is
due to the nearby stable electrostatic modes. In fact in this limit
the tearing mode also becomes electrostatic.

\begin{figure}
\noindent\includegraphics[scale=0.5]{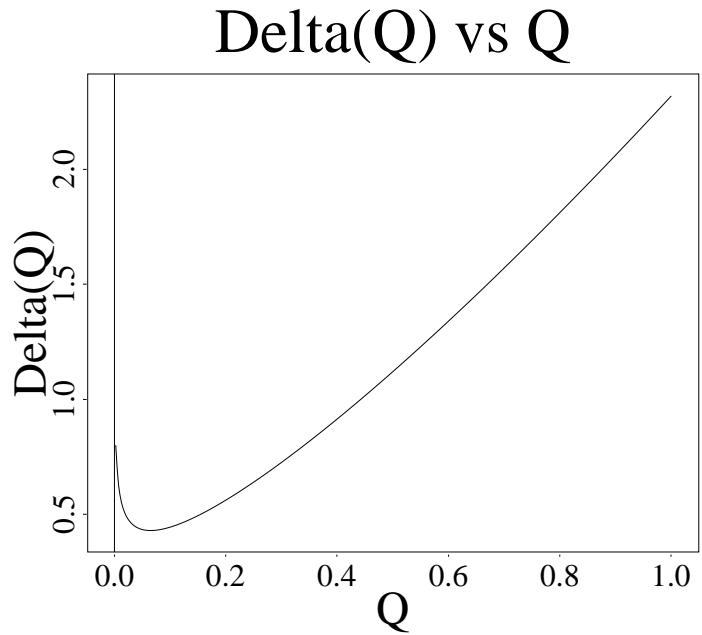}\caption{
Plot of $\Delta(Q)$ in the RI regime with favorable curvature, $g=G_{0}/b_{0}$
negative and small in magnitude (large sound speed.) \label{fig:Delta(Q)-GGJ}}
\end{figure}

The fit in Eq.~(\ref{eq:RI-final-result}) is observed to work well
also in the favorable curvature case ($G_{0}<0$), again for sufficiently
large sound speed (small $|g|$.) Unlike in the unfavorable curvature
case, there is no evidence of electrostatic resistive interchange
poles on the real $Q$ axis. It appears that, as in the case with
$b_{0}=0$ in Sec.~\ref{sub:RI-regime-without-parallel}, these modes
have complex $Q$ for $G_{0}<0$. These results for favorable curvature
show that the behavior observed in Refs.~\cite{GGJ1,GGJ2} holds
qualitatively using only parallel dynamics, without including the
divergence of the $E\times B$ drift and perpendicular particle transport.
That is, we again have $\Delta(Q)=C_{1}Q^{5/4}-C_{2}Q^{-1/4}$ but
with $C_{2}<0$. See Fig.~\ref{fig:Delta(Q)-GGJ}. 
In particular, there are complex roots for $\Delta'<\Delta_{min}=\min(\Delta(Q))\sim|G_{0}|^{5/6}$
and these roots become stable for $\Delta'<\Delta_{crit}\sim|G_{0}|^{5/6}$
\cite{GGJ1,GGJ2}. Numerical results for arbitrary sound speed (arbitrary
values for $g=G_{0}/b_{0}^{2}$) will be shown in a future publication,
with and without perpendicular resistivity and divergence of the $E\times B$
drift.

\section{Viscoresistive (VR) regime with pressure gradient and curvature\label{sec:Viscoresistive-(VR)-regime}}

For the VR regime, we add perpendicular ion viscosity to Eq.~(\ref{eq:vorticity})
but neglect the ion inertia, first without parallel dynamics and then
including parallel dynamics. As in Sec.~\ref{sec:Resistive-inertial-(RI)-regime},
we ignore the divergence of the $E\times B$ drift and the perpendicular
resistivity. The approach is based on that in Sec.~\ref{sec:Resistive-inertial-(RI)-regime}.
As in the RI regime, we keep parallel inertia but \emph{not} parallel
viscosity.

\subsection{\label{sub:Viscoresistive-(VR)-w/o-parallel}VR regime without parallel
dynamics}

In the VR regime, we replace Eq.~(\ref{eq:vorticity}) with 
\begin{equation}
0=iF\nabla_{\perp}^{2}\tilde{\psi}+\frac{2imB_{\theta}^{2}}{B_{0}^{2}r^{2}}\tilde{p}+\mu\nabla_{\perp}^{4}\tilde{\phi},\label{eq:vorticity-2}
\end{equation}
and use Eq.~(\ref{eq:Ohm-adiabatic-bare}). Substituting with the
parallel Ohm's law, Eq.~(\ref{eq:Ohm-adiabatic-bare}a) and the advected
pressure equation, Eq.~(\ref{eq:Ohm-adiabatic-bare}b), we find in
the constant-$\psi$ tearing layer
\begin{equation}
\mu\tilde{\phi}''''+\frac{\alpha^{2}x^{2}}{\eta}-E\tilde{\phi}=-\frac{i\alpha\gamma}{\eta}x\tilde{\psi}_{0},\label{eq:Basic-eq-for-VR}
\end{equation}
where as in the RI regime $E=-2m^{2}B_{\theta}^{2}p'/\gamma B_{0}^{2}r_{t}^{3}=E_{0}/\gamma$.
As usual in the VR regime we find $\delta^{6}=\eta\mu/\alpha^{2}$.
Introducing $\xi=x/\delta$ and $W(\xi)$ so that $\tilde{\phi}=-i\gamma\tilde{\psi}_{0}W(\xi)/\alpha\delta$,
we find
\begin{equation}
\frac{d^{4}W}{d\xi^{4}}+\xi^{2}W-GW=\xi,\label{eq:Basic-VR-eq-for-W}
\end{equation}
where in this regime we have $G=E\delta^{4}/\mu=E_{0}\delta^{4}/\gamma\mu$.
Integration of the parallel Ohm's law yields $\Delta(\gamma)=[\tilde{\psi}']_{0}/\tilde{\psi}_{0}$
or
\begin{equation}
\Delta(\gamma)=\frac{\gamma\delta}{\eta}\Delta_{s}(Q)=Q\Delta_{s}(Q),\,\,\,\,\,\,\Delta_{s}(Q)=\int_{-\infty}^{\infty}\left(1-\xi W\right)d\xi,\label{eq:VR-diepersion-relation-final}
\end{equation}
where $\gamma/\gamma_{vr}=Q$ with $\gamma_{vr}=1/\tau_{vr}=\eta/\delta$.
As usual one matches with $\Delta'$ from the outer region. Noting
that $\delta$ is independent of $\gamma$ in this regime, leading
to 
\begin{equation}
G=\frac{G_{0}}{Q}.\label{eq:VR-G-in-terms-of-G0}
\end{equation}
For $p'<0$ in cylindrical geometry, we have $G_{0}>0$, and in toroidal
geometry we have $G_{0}\rightarrow(1-q(r_{t})^{2})G_{0}$. For magnetic
Prandtl number $Pr=\mu/\eta=1$, taking $\delta\sim\eta^{1/3}=\epsilon$,
we have $\gamma\sim\epsilon^{2}$. Ordering $E_{0}\sim\beta\sim\epsilon$
leads to $G_{0}\sim1$. 

As in the RI regime in the last section, there are homogeneous modes,
localized modes satisfying the homogeneous form of Eq.~(\ref{eq:Basic-VR-eq-for-W}).
This equation has an infinite number of real eigenvalues $Q_{n}>0$
for $G_{0}>0$, the localized electrostatic resistive interchange
modes. In Fourier space, the homogeneous form of Eq.~(\ref{eq:Basic-VR-eq-for-W})
takes the Schr\"odinger form $d^{2}\hat{W}/dk^{2}+(G-k^{4})\hat{W}=0$
(quartic oscillator) with $G=G_{0}/Q$. The eigenvalues $G$ are the
energy levels for the quartic oscillator, having%
\footnote{This is most easily seen by computing the action $\oint(E-k^{4})^{1/2}dk\propto n$.%
} $G\propto n^{4/3}$ for large $n$, leading to 
\begin{equation}
Q_{n}\sim\frac{G_{0}}{n^{4/3}},\label{eq:VR-es-Modes}
\end{equation}
with $\gamma_{es}=\gamma_{vr}Q_{n}$. Corresponding to these, the
quantity $\Delta_{s}(Q)$ has poles at $Q_{n}$. Note that, as in
the RI regime (Eq.~(\ref{eq:Hermite})), the eigenvalues (poles of
$\Delta_{s}(Q)$) $Q_{n}$ have an accumulation point at $Q=0$. There
are no eigenvalues (poles) on the negative real axis.

Equation (\ref{eq:Basic-VR-eq-for-W}) has the symmetry $G_{0}\rightarrow\lambda G_{0}$,
$Q\rightarrow\lambda Q$, and for this regime these scalings hold
for either sign of $\lambda$. The quantity $\Delta_{s}$ is unchanged
under this transformation, and we have $\Delta(Q)=Q\Delta_{s}(Q)\rightarrow\lambda\Delta(Q)$.
Therefore, using Eq.~(\ref{eq:VR-es-Modes}) we see that for $G_{0}<0$
there is an infinite number of negative real eigenvalues also with
an accumulation point at $Q=0$, $Q_{n}\sim G_{0}/n^{4/3}$. These
are stable localized electrostatic resistive interchange modes, on
the negative real axis for $G_{0}<0$ in this regime. For electromagnetic
tearing modes coupled to the outer region, the quantity $G=G_{0}/Q$
is inversely related to the growth rate relative to the growth rate
for the electrostatic modes.

Away from the poles with $G_{0}>0$ , increasing $G_{0}$ reduces
$\Delta_{s}$ and therefore destabilizes. If $G_{0}<0$ (either $p'>0$
or with the toroidal factor $1-q(r_{t})^{2}$), there are no poles
on the real $Q$ axis and $\Delta_{s}$ is increased. Therefore the
growth rate for the mode is reduced.

\subsection{VR regime with parallel dynamics\label{sub:VR-regime-with-parallel}}

With pressure gradient, curvature and parallel dynamics, but without
perpendicular compression or particle transport, the substitution
in Eq.~(\ref{eq:Replacement}) is again valid and the inner region
equation for the streamfunction takes the form

\begin{equation}
\frac{d^{4}W}{d\xi^{4}}+\xi^{2}W-\frac{GQ^{2}}{Q^{2}+b^{2}\xi^{2}}W=\left(1+\frac{Gb^{2}}{Q^{2}+b^{2}\xi^{2}}\right)\xi,\label{eq:W-equation-in-VR}
\end{equation}
where $r-r_{t}=\delta\xi$ with $\delta=(\eta\mu/\alpha^{2})^{1/6}$,
$Q=\gamma\tau_{vr}=\gamma/\gamma_{vr}$, $G\propto D_{s}$ is defined
as in Sec.~\ref{sub:Viscoresistive-(VR)-w/o-parallel}, and $b^{2}=\alpha^{2}\delta^{2}c_{s}^{2}/\gamma_{vr}^{2}B_{0}^{2}$.
The matching condition is again $\Delta'=Q\Delta_{s}(Q)$. Equation
(\ref{eq:W-equation-in-VR}) has 
\begin{equation}
G=\frac{G_{0}}{Q},\,\,\,\,\, b=b_{0},\label{eq:b-in-terms-of-b0-VR}
\end{equation}
where $G_{0}=E_{0}\delta^{4}/\gamma_{vr}\mu$. The fact that $b$
is independent of $Q$ is due to the independence of $\delta$ on
$\gamma$ in the VR regime. 

Equation (\ref{eq:W-equation-in-VR}) has the symmetry $G_{0}\rightarrow\lambda G_{0},\,\, b_{0}\rightarrow\lambda b_{0},\,\, Q\rightarrow\lambda Q$,
so the invariants are $G_{0}/Q$ and $b_{0}/Q$, or equivalently $G=G_{0}/Q$
and $g=G_{0}/b_{0}$. As in Sec.~\ref{sub:Viscoresistive-(VR)-w/o-parallel},
$G=G_{0}/Q$ is inversely related to the growth rate relative to that
of the electrostatic modes. The quantity $(b_{0}/Q)^{2}\sim k_{||}^{2}c_{s}^{2}/\gamma^{2}$
measures the stabilizing influence of sound wave propagation in the
layer ($\xi^{2}\sim1$), the third term in Eq.~(\ref{eq:W-equation-in-VR}).
The quantity $Gb^{2}/Q^{2}=\left(G_{0}/Q\right)\left(b_{0}/Q\right)^{2}$
measures the importance of the last term in Eq.~(\ref{eq:W-equation-in-VR}),
i.e.~the $p'\tilde{\psi}_{0}$ term in Eq.~(\ref{eq:vParallelEq}).
The quantity $G_{0}/b_{0}\sim\gamma_{es}/k_{||}c_{s}$ measures the
growth rate of the electrostatic modes relative to the sound propagation
term. As in the VR regime without parallel dynamics, this symmetry
extends to $\lambda<0$, i.e.~$Q\rightarrow-Q,\,\, G_{0}\rightarrow-G_{0}$
(and trivially to $b_{0}\rightarrow-b_{0}$.) The quantity $\Delta_{s}$
is unchanged by this transformation again, but $\Delta(Q)\rightarrow\lambda\Delta(Q)$,
i.e.~$\Delta$ changes sign with $G_{0}$. Thus the results for unfavorable
curvature ($G_{0}>0$) can be applied directly to the favorable curvature
($G_{0}<0$) case. As in the RI regime, $\Delta_{s}$ depends only
on the group invariants, here $G=G_{0}/Q$ and $g=G_{0}/b_{0}$.

\begin{figure}
\noindent\includegraphics[angle=270,scale=0.5]{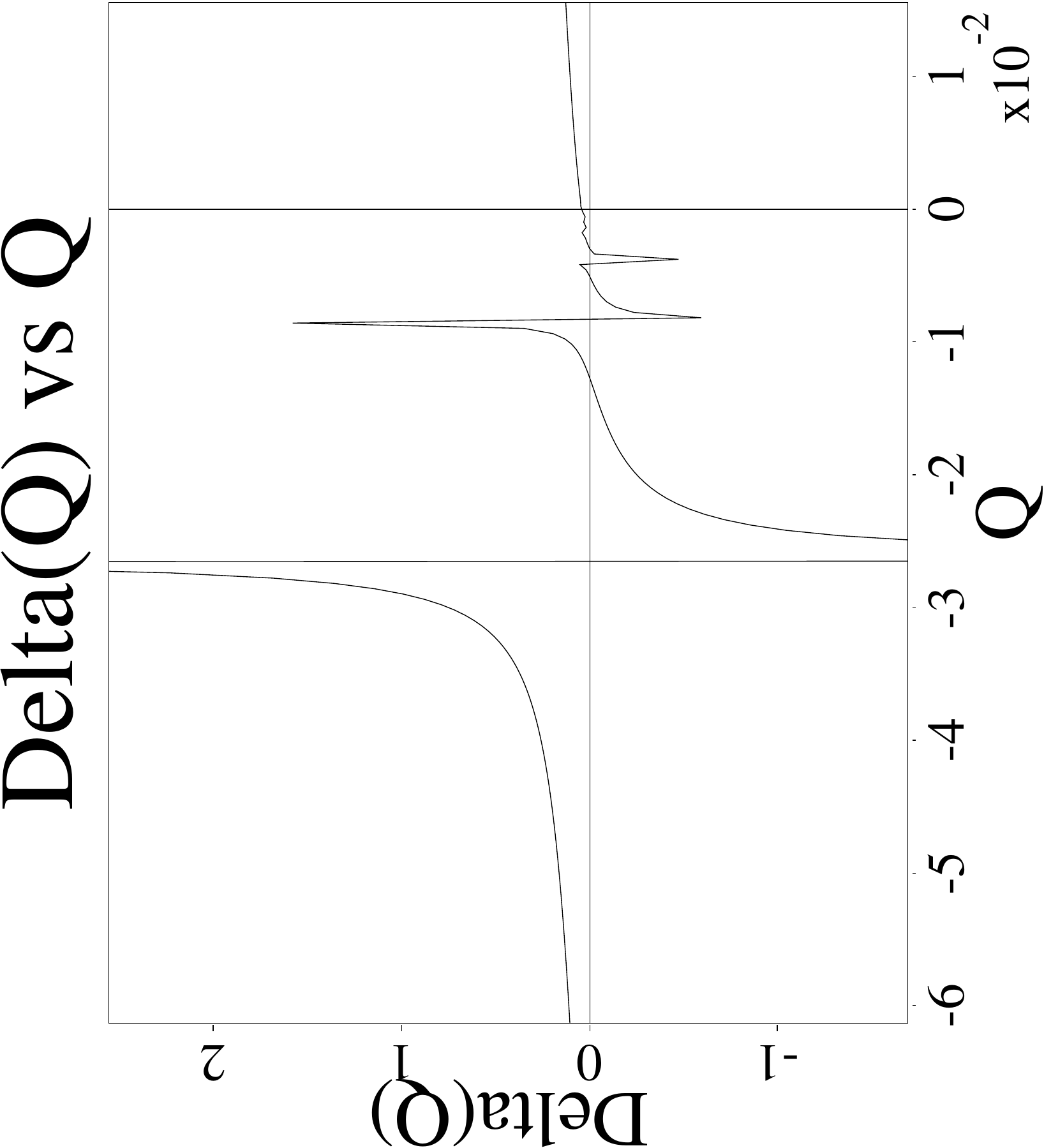}\caption{
Plot of $\Delta(Q)$ in the VR regime with favorable curvature for
low sound speed, (small $b_{0}$, large $g=G_{0}/b_{0}$), showing
three poles on the negative real axis, corresponding to three stable
electrostatic interchange modes. \label{fig:Delta_S(Q)-VR-3-poles}}
\end{figure}

We will describe the results first in terms of favorable curvature
($G_{0}<0)$. For low sound speed (large values of $|G_{0}/b_{0}|=|g|\sim|p'|/\sqrt{p}$),
the plot of $\Delta_{s}(Q)$ on the real axis shows poles $Q=Q_{n}<0$
(with $n$ odd again.) Again, these correspond to stable electrostatic
resistive interchanges, solutions to the homogeneous form of Eq.~(\ref{eq:W-equation-in-VR}).
See Fig.~\ref{fig:Delta_S(Q)-VR-3-poles}.

As the sound speed increases, i.e.~as $|g|$ decreases, all but the
last two of these modes go to $Q=0$ and become complex. See Fig.~\ref{fig:Delta_S(Q)-VR-3-poles}.
\begin{figure}
\noindent\includegraphics[angle=270,scale=0.5]{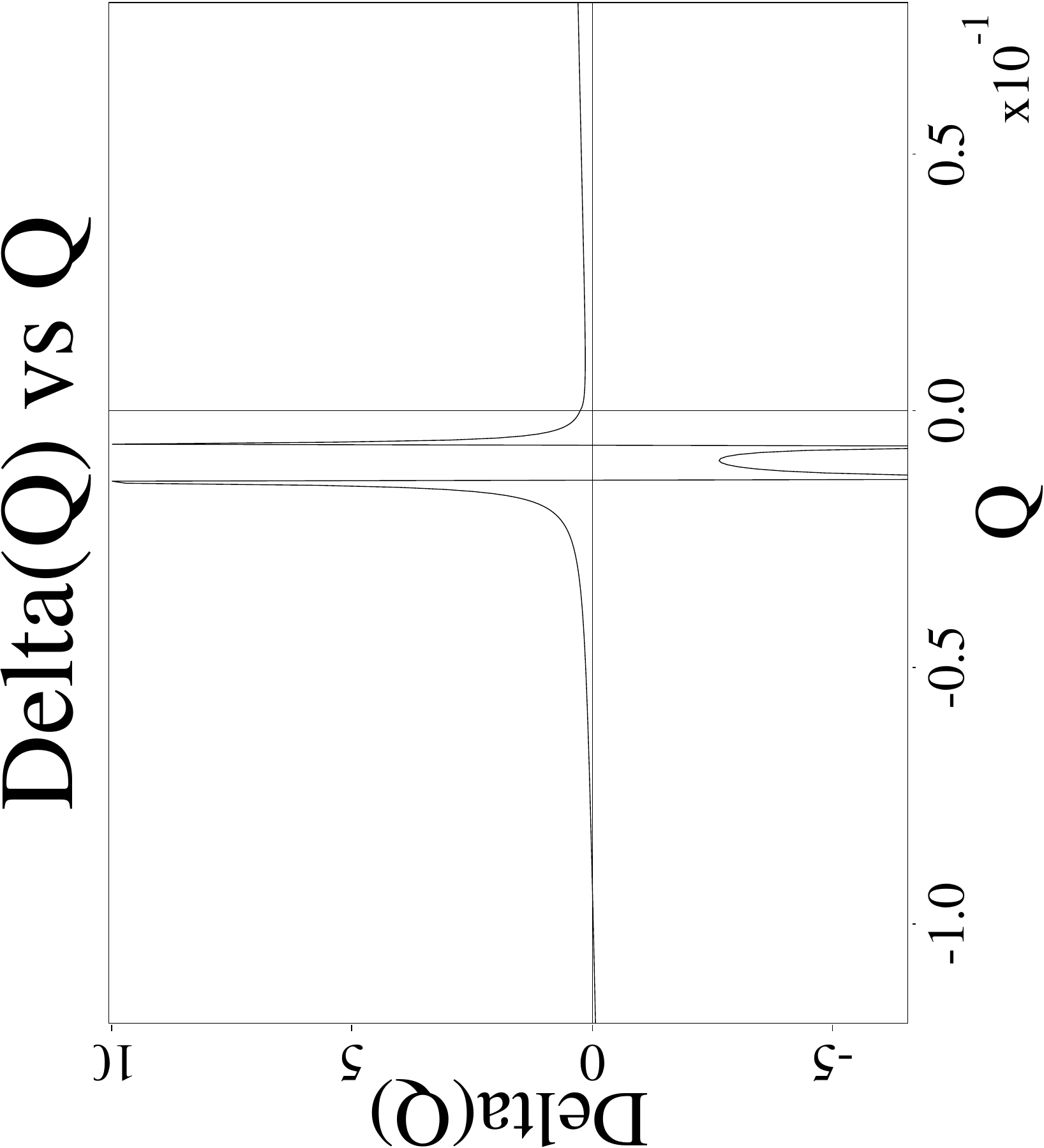}
\caption{Plot of $\Delta(Q)$ in the VR regime with favorable curvature for
higher sound speed, $G_{0}=-0.1,\, b_{0}=0.0102$ ($g=G_{0}/b_{0}=-9.8$),
showing two remaining poles on the real axis just before they coalesce
and go into the complex plane. \label{fig:Delta(Q)-2-close-poles}}
\end{figure}
The last two poles on the negative real axis (the last two interchange
modes with $W$ odd in $\xi$) become complex at $g=-1.2$ for negative
real part of $Q$, leaving a nonmonotonic curve of $\Delta(Q)=Q\Delta_{s}(Q)$
on the real axis, as shown in Fig.~\ref{fig:Delta(Q)-nonmonoton}.
This behavior is related to the proximity of the complex roots corresponding
to electrostatic modes. 
\begin{figure}
\noindent\includegraphics[width=4in]{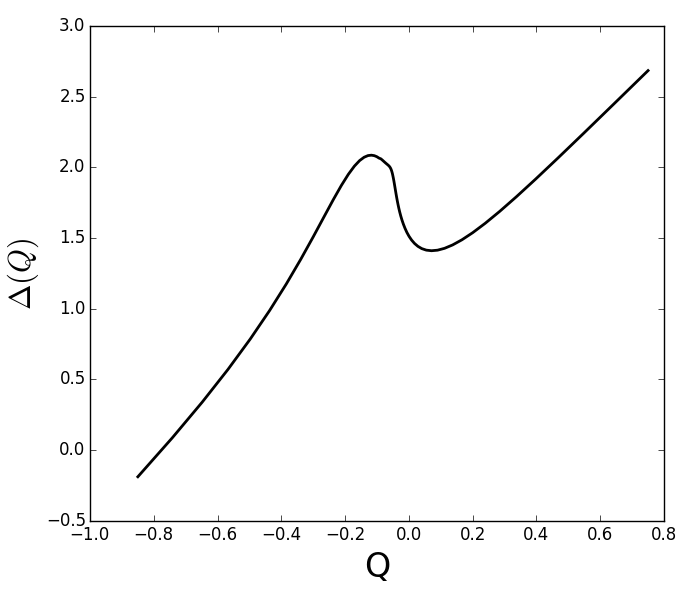} \caption{
Plot of $\Delta(Q)$ in the VR regime with favorable curvature for
higher sound speed, $G_{0}=-0.8,\, b_{0}=0.18$ ($g=G_{0}/b_{0}=-0.44$),
showing non-monotonic behavior that arises just after the poles in
Fig.~\ref{fig:Delta(Q)-2-close-poles} coalesce. Unstable complex
conjugate roots occur for $\Delta'$ just below the local minimum
of $\Delta(Q)$. \label{fig:Delta(Q)-nonmonoton}}
\end{figure}
For $\Delta'$ above the relative minimum $\Delta=\Delta_{min}$ in
Fig.~\ref{fig:Delta(Q)-nonmonoton}, there are two unstable modes
with real $Q$. Just below $\Delta_{min}$, there are complex tearing
mode (electromagnetic) roots, and the locus of these roots is similar
to that for the RI regime, shown in Fig.~\ref{fig:VR-locus(T!)}.
In particular, there is a value $\Delta_{crit}>0$, $\Delta_{crit}<\Delta_{min}$
for which the modes with complex roots become stable. 
\begin{figure}
\noindent\includegraphics[scale=0.5]{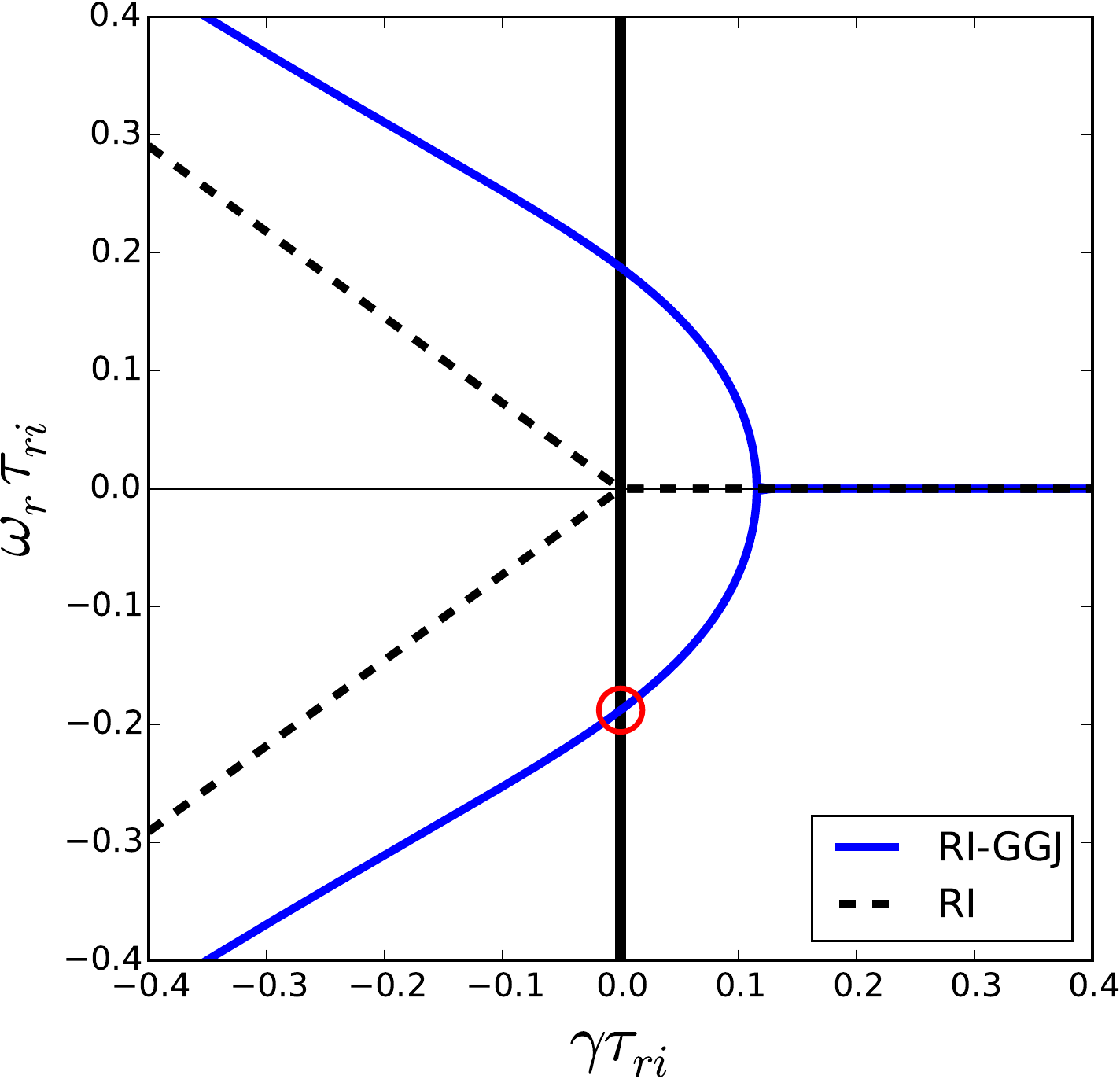}\caption{
Locus of complex roots in the RI regime with favorable curvature,
as $\Delta'$ decreases (blue). The black dashed curve is the locus
of RI roots for $G_{0}=0$. The marginally stable root encircled in
red, with $\omega_{r}<0$, can be Doppler shifted to zero by $v=-\omega_{r}/k$.\textbf{\textcolor{red}{{}
}}\label{fig:VR-locus(T!)}}
\end{figure}
The curve $\Delta(Q)$ is nonmonotonic for the wide range $-9.6<g<-1.2$.
Below $g=-9.6$, $\Delta(Q)$ becomes monotonic, and becomes a straight
line as $g\rightarrow-\infty$. This straight line behavior was observed
in Ref.~\cite{CGJ} for very large sound speed and both signs of
$D_{s}\propto G_{0}$.\textcolor{red}{{} }From these results it is clear
that complex roots for $\Delta'<\Delta_{min}$ and stability for $\Delta'<\Delta_{crit}$
with $\Delta_{crit}>0$ occurs, as in the RI regime with parallel
dynamics. We conclude that the Glasser effect occurs in the VR regime
as well, over a wide range of parameters.

For unfavorable curvature ($G_{0}>0$), the symmetry $G_{0}\rightarrow-G_{0}$,
$Q\rightarrow-Q$ shows that similar poles occur, but they correspond
to growing modes when the roots for $G_{0}<0$ are damped. The last
of these poles become complex at $g=1.2$. Past this value, non-monotonic
behavior of $\Delta(Q)$, similar to that with favorable curvature
and also due to the proximity of the complex electrostatic roots,
occurs up to $g=9.6$.  However, this non-monotonic behavior is less
important for unfavorable curvature, since the electrostatic modes
are unstable.

Numerical results for arbitrary balues of $g=G_{0}/b_{0}$ will be
presented in a future publication.

\section{Resistive wall tearing modes and locking\textcolor{red}{\label{sec:Resistive-wall-tearing}}}

In this section we present results on error field locking and resistive
wall tearing modes, for tearing layers with real frequencies in the
plasma frame. For both situations the main effect occurs when the
Doppler shift due to the $E\times B$ rotation brings the mode close
to zero frequency in the laboratory frame.

\subsection{Resonant field amplification and locking torques\label{sub:Resonant-field-amplification}}

An important consequence of the presence of tearing modes with real
frequencies in the plasma frame is related to error field penetration
and error field amplification. Recently, a study has been made of
the Maxwell torque on a rotating toroidal plasma caused by an error
field at rest in the laboratory frame. It was shown \cite{FinnColeBrennan}
that the Maxwell torque on the plasma, applied across the tearing
layer, is zero at the velocity for which the frequency of the spontaneous
tearing mode in the laboratory frame $\omega_{r}+kv$ is zero, i.e.~for
$v=-\omega_{r}/k$. (We write the Doppler shift as $\mathbf{k\cdot v}=kv$
because, as discussed below, the poloidal rotation is small. Also,
the torque is only exactly zero at $v=-\omega_{r}/k$ in the limit
$\gamma\rightarrow0-$, where $\gamma$ is the growth rate of the
spontaneous tearing mode.) We know that spontaneous modes have complex
conjugate growth rates $\gamma_{c}=\gamma\mp i\omega_{r}$. For such
cases the mode with $\omega_{r}<0$ (the \emph{`backward wave'}) has
frequency in the laboratory frame $\omega_{r}+kv$ equal to zero for
positive $v=-\omega_{r}/k$. The existence of complex conjugate roots
implies that the other `backward wave', for $\omega_{r}>0$, has zero
frequency in the laboratory frame for $v=-\omega_{r}/k$, which is
negative. A related phenomenon is error field amplification or resonant
field amplification, i.e.~amplification of the reconnected flux $\tilde{\psi}$
at the tearing layer relative to the magnitude of the error field.
This quantity is large when the spontaneous tearing mode is weakly
damped, but is maximized when the above condition $v=\pm\omega_{r}/k$
holds. (Again, this is the exact velocity of the peak in reconnected
flux only in the limit $\gamma\rightarrow0-$.) These conclusions
imply that the reconnected flux is symmetric with respect to $v$.
As we will discuss later, other tearing modes do not have complex
conjugate roots, so that this symmetry is not present.

As in Ref.~\cite{FinnColeBrennan}, we expand $\psi(r)=\alpha_{1}\phi_{1}(r)+\alpha_{2}\phi_{2}(r)$,
where $\phi_{1}(r_{w})=0$ and $\phi_{2}(r_{t})=0$; here $\phi_{1}$
is the flux for the ideal wall tearing mode and $\phi_{2}$ is the
flux to the right of the mode rational surface at $r_{t}$ allowing
for the error field at the wall, at $r_{w}$. Using the constant-$\psi$
approximation, the reconnected flux ar $r=r_{t}$ is found to be
\begin{equation}
\psi(r_{t})=\frac{l_{21}}{\Delta(ikv)-\Delta_{1}}\tilde{\psi}(r_{w}),\label{reconnected-flux}
\end{equation}
where $l_{21}=\phi_{2}'(r_{t}+)$, $\tilde{\psi}(r_{w})$ is the error
field, $\Delta_{1}=[\tilde{\psi}']_{r_{t}}$ and $\Delta(\gamma)$,
used here with $\gamma\rightarrow ikv$, is the dispersion quantity
discussed in Secs.~\ref{sec:Resistive-inertial-(RI)-regime} and
\ref{sec:Viscoresistive-(VR)-regime}, and $\Omega=\mathbf{k\cdot v}$
is the Doppler shift. The denominator $\Delta-\Delta_{1}$ vanishes
when the ideal wall (spontaneous) tearing mode dispersion relation
is satisfied. The reconnected flux has peaks near marginal stability
$\gamma\lesssim0$ and, for spontaneous modes with real frequencies,
when $kv=\pm\omega$. Except for the inclusion of pressure-curvature
terms in $\Delta(Q)$, these layer calculations can be perfomed in
slab geometry.

The Maxwell force on the tearing layer $\int\tilde{j}_{||}\tilde{B}_{x}dV$
is given by \cite{ColeFinnHegnaTerry}
\[
F_{m}=-\frac{k_{x}}{2}\text{Im}(\Delta(ikv))|\tilde{\psi}(r_{t})|^{2}
\]
\begin{equation}
=-\frac{k_{x}}{2}\text{Im}(\Delta(ikv))\frac{l_{21}^{2}}{|\Delta(ikv)-\Delta_{1}|^{2}}|\tilde{\psi}(r_{w})|^{2}.\label{eq:General-force-formula}
\end{equation}
In toroidal geometry, the force in the poloidal direction is dominated
by magnetic pumping, so that, as mentioned above, this leads to essentially
zero poloidal rotation, giving $\mathbf{k\cdot v}=kv$. The toroidal
force is the quantity in Eq.~(\ref{eq:General-force-formula}), as
usual reduced by the factor $-r_{t}/q(r_{t})R$, and the associated
torque is $N_{m}=RF_{m}$, where $R$ is the major radius.

The reconnected flux and Maxwell torque for real frequency tearing
modes in the RI regime with pressure and field line curvature $D\sim p'\kappa$
(Glasser effect) were studied in Ref.~\cite{FinnColeBrennan}. In
Fig.~8a we show the reconnected flux for the VR regime with $D<0$.
The quantity $Q_{0}$ is defined below. In Fig.~8b we show the Maxwell
force (or torque) and the viscous torque, due to an external source
of momentum and plasma viscosity\cite{FinnColeBrennan}. The qualitative
behavior is as in Ref.~\cite{FinnColeBrennan}. Specifically, there
are three ranges of parameters: one for which there is only a locked
state, one for which there is only an unlocked state, and a third
in which there are three possible states, with only the locked and
unlocked states stable. Because the torque $N_{m}=RF_{m}$ increases
sharply, the locked state is just to the right of $v=-\omega_{r}/k$.
Furthermore, the velocity of the locked state asymptotes to this value
as the error field $\tilde{\psi}(r_{w})$ goes to infinity, i.e.~as
$N_{m}\sim|\tilde{\psi}(r_{w})|^{2}\rightarrow\infty$. Also, as in
Ref.~\cite{FinnColeBrennan}, we find that there are two other possible
states with negative velocities, and only the one with more negative
velocity is stable. Notice the symmetry for $v\rightarrow-v$ in the
reconnected flux and the antisymmetry in the torque. We will return
to the issue of other regimes without these symmetry properties.

For computing torques for slightly damped spontaneous modes, there
is another complication: the presence of the denominators $Q^{2}+b^{2}\xi^{2}\sim-\omega^{2}+k_{||}^{2}c_{s}^{2}$
(sound wave propagators) in Eq.~(\ref{eq:W-equation-in-VR}) lead
to sound wave resonances and continuum-like effects because the equations
for $\tilde{p}$ and $\tilde{v}_{||}$, Eqs.~(\ref{eq:AdiabatWithParallel})
and (\ref{eq:vParallelEq}), have no dissipation. This resonance is
on the imaginary $Q$ axis so that slightly damped modes are close
to this continuum. This continuum can be moved to the left in the
complex $Q$ plane by $Q\rightarrow Q+Q_{0}$ in the sound wave propagators,
i.e.~by including drag terms either Eq.~(\ref{eq:AdiabatWithParallel})
or Eq.~(\ref{eq:vParallelEq}) or both. The inclusion of perpendicular
resistivity in Eq.~(\ref{eq:vParallelEq}) increases the order of
the equations and has the much more profound effect of removing this
sound continuum. Details will be presented in a future publication.
Interestingly, in the RI regime with pressure gradient and curvature,
this continuum (for $Q_{0}=0$) is along the rays $Q\propto e^{\pm2\pi i/3}$
with $Q_{r}<0$ (except for $Q_{i}=0$, $Q_{r}=0$). This property
follows from the fact that $\omega^{2}-k_{||}^{2}c_{s}^{2}=0$ has
$Q^{2}\sim-b_{0}^{2}Q^{1/2}$ for $\xi^{2}\sim1$.

\begin{figure}
\noindent\includegraphics[clip,width=3in]{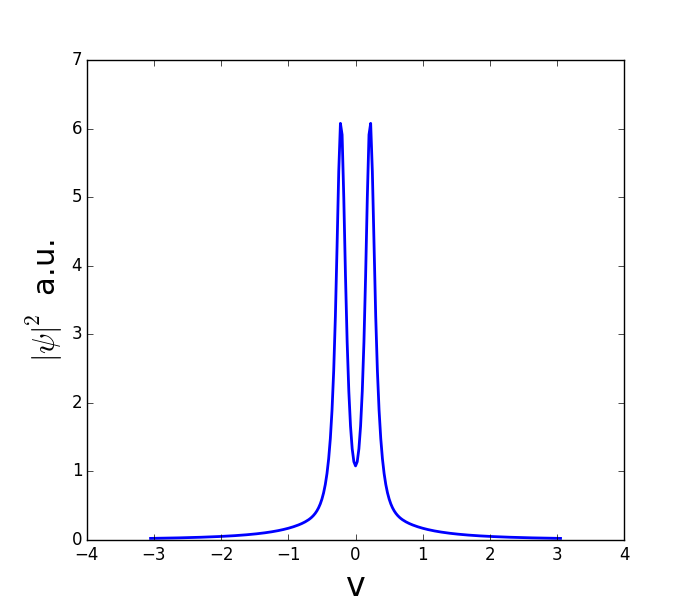} \begin{raggedright}
\protect\protect\includegraphics[width=3in]{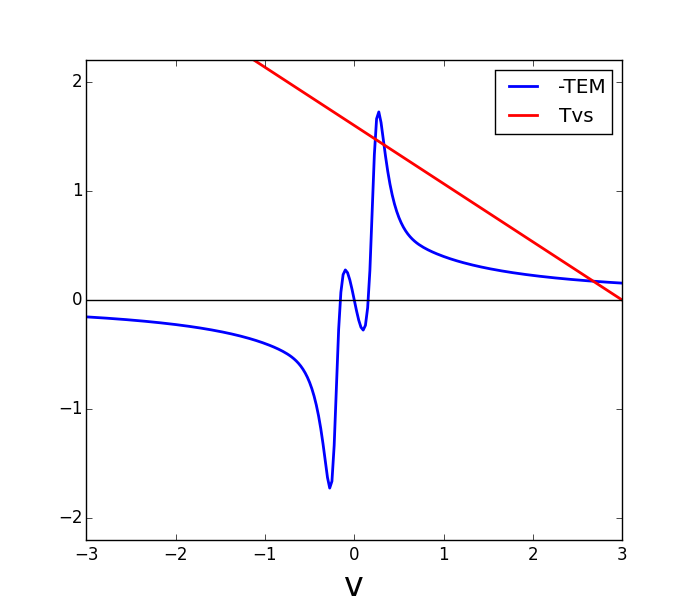}

\par\end{raggedright}

\noindent \caption{Reconnected flux $|\psi|^{2}$ (a) and torques
$N_{m}$ (labeled TEM) (b) for the VR Glasser regime, for a case with
a slightly damped spontaneous tearing mode, with $G_{0}=-0.8,\, b_{0}=0.18$,
and for drag coefficient $Q_{0}=0.05$. The intersecting viscous torque
$T_{vs}$ is also shown in (b). The peaks in $|\psi|^{2}$ coincide
approximately with $v=\pm\omega_{r}/k$, as do the zeroes of the torque
$N_{m}$ away from $v=0$. The leftmost intersection with the viscous
torque line is just above $v=\omega_{r}/k$. \label{fig:ReconFluxTorques}}
\end{figure}

The results in the RI regime and in the VR regime, both with $D<0$,
have the property discussed above that modes with complex growth rates
occur in complex conjugate pairs $\gamma_{c}=\gamma\pm i\omega_{r}$,
leading to the symmetries apparent in Fig.~8. Such symmetries may
not hold in other regimes with real frequencies, e.g.~when $\omega_{r}$
is due to electron diamagnetism, so that spontaneous modes have $\omega_{r}\sim\omega_{*}$,
the diamagnetic frequency (drift-tearing modes.)

\subsection{Resistive wall tearing modes with real frequency tearing layers\label{sub:Resistive-wall-tearing}}

A closely related issue is that of resistive wall tearing modes, in
which the tearing layers have real frequencies. First, let us review
the situation without real frequencies. In the case of VR layers without
pressure, the ideal wall tearing modes have real $\gamma$ for stable
as well as unstable modes, as discussed in Ref.~\cite{FinnGerwinModeCoupling}.
This is easily seen by an analysis as in Sec.~\ref{sub:Resonant-field-amplification}.
We use $\gamma\tau_{vr}\tilde{\psi}(r_{t})=[\tilde{\psi}']_{r_{t}}$,
i.e.~the constant-$\psi$ VR relation for the tearing layer and $\gamma\tau_{w}\tilde{\psi}(r_{w})=[\tilde{\psi}']_{r_{w}}$,
the thin-wall (constant-$\psi$) relation for the resistive wall.
By methods as in Sec.~\ref{sub:Resonant-field-amplification} we
find
\begin{equation}
\left[\begin{array}{cc}
\Delta_{1}-\gamma\tau_{vr} & l_{21}\\
l_{12} & \Delta_{2}-\gamma\tau_{w}
\end{array}\right]\left(\begin{array}{c}
\alpha_{1}\\
\alpha_{2}
\end{array}\right)=\left[\begin{array}{cc}
D_{1}(\gamma) & l_{21}\\
l_{12} & D_{2}(\gamma)
\end{array}\right]\left(\begin{array}{c}
\alpha_{1}\\
\alpha_{2}
\end{array}\right)=0,\label{eq:VR-RWM}
\end{equation}
where again $l_{21}=\phi_{2}'(r_{t}+)$, $\Delta_{1}=[\phi_{1}']_{r_{t}}$.
We have used $D_{1}(\gamma)=\Delta_{1}-\Delta(\gamma)=\Delta_{1}-\gamma\tau_{vr}$,
The quantity $\Delta_{1}$ determines the stability of the ideal wall
tearing mode ($\tau_{w}=\infty$); it depends on the current profile
and also on the pressure profile. We also have $l_{12}=-\phi_{1}'(r_{w}-)$
and $\Delta_{2}=[\phi_{2}']_{r_{w}}$. Both $l_{21}$ and $l_{12}$
are positive. The quantity $\Delta_{2}$ depends on the location of
a second, conducting, wall at $r_{c}>r_{w}$. The possibility $r_{c}=\infty$
is allowed, and $\Delta_{2}$ determines the stability of the ideal
plasma resistive wall mode ($\tau_{vr}=\infty$).%
\footnote{This equation can also be put into the conventional eigenvalue form
$(\Delta_{1}/\tau_{vr})\alpha_{1}+(l_{21}/\tau_{vr})\alpha_{2}=\gamma\alpha_{1}$,~~$(l_{12}/\tau_{w})\alpha_{1}+(\Delta_{2}/\tau_{w})\alpha_{2}=\gamma\alpha_{2}$.
Then, letting $\alpha_{2}\rightarrow\beta\alpha_{2}$, and using $l_{21}>0,\,\, l_{12}>0$
we find that the new matrix corresponding to Eq.~(\ref{eq:VR-RWM})
can be put into a symmetric form.%
} 

Plasma $E\times B$ rotation is included by letting $\gamma\rightarrow\gamma+i\Omega$,
where $\Omega=kv(r_{t})$. We can compute the critical value for $\Delta_{1}$
(critical value of $\beta$, where the pressure gradient-curvature
drive comes from the ideal outer region) for which the more unstable
mode is marginally stable. As discussed in Ref.~\cite{FinnGerwinModeCoupling},
this critical value always increases with $\Omega$ for the VR model.
The interpretation of this result is that the two uncoupled modes
in the presence of rotation, with $\gamma=\Delta_{1}/\tau_{vr}-i\Omega$
and $\gamma=\Delta_{2}/\tau_{w}$, are closest in the complex plane
for $\Omega=0$, and therefore are coupled most strongly there. The
Doppler shift, for $\Omega$ positive or negative, only suppresses
the flux from penetrating the wall. Thus, rotation is always stabilizing
for the VR model.

In the RI regime, again without pressure, these results are modified
by $\gamma\tau_{vr}\rightarrow(\gamma\tau_{ri})^{5/4}$. As also discussed
in Ref.~\cite{FinnGerwinModeCoupling}, the critical $\Delta_{1}$
(critical $\beta$) decreases slightly for small $\Omega$, followed
by an increase. This is also explained by the mode coupling picture:
the uncoupled modes, with $\gamma=\Delta_{1}^{4/5}/\tau_{ri}-i\Omega$
and $\gamma=\Delta_{2}/\tau_{w}$, are closest for negative $\Delta_{1}$
when $\Omega$ equals $-\omega_{r}$, the imaginary part of $\Delta_{1}^{4/5}/\tau_{ri}$,
namely $=\pm|\Delta_{1}|^{4/5}\sin(4\pi i/5)\tau_{ri}$. Therefore
the strongest coupling is for either of these values of $\Omega$,
and further increases in $|\Omega|$ are stabilizing. Again, the strongest
destabilizing effect is for that value of $\Omega$ for which one
of the real frequencies is Doppler shifted to have zero phase velocity
in the laboratory frame. However, as discussed in Ref.~\cite{FinnGerwinModeCoupling},
this effect is very weak for the RI regime without pressure, because
$|\omega_{r}|$ becomes appreciable only when the growth rate $|\Delta_{1}|^{4/5}\cos(4\pi i/5)\tau_{ri}$
becomes strongly negative. (On the other hand, it was found \cite{FinnGerwinModeCoupling}
that \emph{ideal plasma} resistive wall modes, with purely real frequencies
for stable modes, can be strongly destabilized over a wide range of
rotation values. \textcolor{red}{})

For tearing mode regimes in which there is a nonzero value of real
frequency near marginal stability, as discussed in Secs.~\ref{sub:RI-regime-with-parallel}
and \ref{sub:VR-regime-with-parallel}, the effect of rotation on
the resistive wall tearing mode is much more significant that it is
in the RI regime in Ref.~\cite{FinnGerwinModeCoupling}. This effect
is most pronounced if the ideal wall tearing mode is very weakly damped.
 These results are also explained by the mode coupling picture of
Ref.~\cite{FinnGerwinModeCoupling}. %
\footnote{As in error field amplification, this matching is exact only in the
limit $\gamma\rightarrow0-$, where $\gamma$ is the growth rate of
the ideal wall tearing mode.%
} 

We have found similar results for resistive wall tearing modes with
layers in the VR regime with pressure. These effects are due to the
real frequencies (Glasser effect) found in this regime, as shown in
Sec.~\ref{sub:VR-regime-with-parallel}. These results are qualitatively
different from those in Ref.~\cite{FinnGerwinModeCoupling}, in which
the resistive wall tearing mode in the VR regime without pressure
was shown to be stabilized for any value of $\Omega$. 

There is an interesting application to double tearing modes, of interest
in reversed shear profiles. If such a mode has both layers in the
VR regime, the mode is described by Eq.~(\ref{eq:VR-RWM}), with
$\tau_{vr}\rightarrow\tau_{vr1}$ and $\tau_{w}\rightarrow\tau_{vr2}$.
That is, a second tearing layer in the VR regime acts like a resistive
wall. In this formalism, the two single tearing modes (e.g.~in the
VR regime), with $\gamma\tau_{vr1}=\Delta_{1}$ ($\alpha_{2}=0$,
e.g.~$\tau_{vr2}=\infty$) and $\gamma\tau_{vr2}=\Delta_{2}$ ($\alpha_{1}=0$,
e.g.~$\tau_{vr1}=\infty$) are coupled by $l_{12}l_{21}$, giving
a double tearing mode which is more unstable than either single tearing
mode. The formulation of Eq.~(\ref{eq:VR-RWM}) shows that relative
rotation of the two resonant surfaces only stabilizes this double
tearing mode with layers in the VR regime. If, on the other hand one
or both layes are in regimes with real frequencies, $\gamma\tau_{vr1}\rightarrow D_{1}(\gamma)$
or $\gamma\tau_{vr2}\rightarrow D_{2}(\gamma)$, then the situation
becomes identical to that studied in this section for resistive wall
modes. A finite value of the relative rotation of the two layers can
destabilize, leading to a maximum growth rate very near where $\omega_{r1}+kv(r_{1})$
equals $\omega_{r2}+kv(r_{2})$. That is, sheared rotation (i.e.~different
rotation rates at the two rational surfaces) can \emph{destabilize}
double tearing modes. Finally, a similar effect can occur in coupling
of the $m$ and $m\pm1$ poloidal Fourier harmonics in toroidal geometry.
These points will be expanded upon in a future publication.

\section{Summary and discussion\label{sec:Conclusions}}

We have first re-analyzed the tearing mode in the RI regime in the
presence of pressure gradient and favorable or unfavorable field line
curvature in the tearing layer, with parallel dynamics. For unfavorable
curvature the dispersion relation $\Delta'=\Delta(Q)$ on the real
$Q$ axis ($Q$ is the dimensionless, possibly complex, growth rate)
has poles, corresponding to unstable electrostatic resistive interchanges.
As the sound speed increases, these poles disappear from the real
$Q$ axis at $Q=0$. For large enough sound speed the dispersion relation
has the known form \cite{GGJ1,GGJ2,FinnManheimer} $\Delta(Q)=C_{1}Q^{5/4}-C_{2}/Q^{1/4}$,
with $C_{2}$ proportional to the pressure gradient. The behavior
$\Delta(Q)\propto-Q^{-1/4}$ for unfavorable curvature as $Q\rightarrow0$
appears to be a remnant of the last electrostatic pole that disappears
into $Q=0$. For favorable curvature (change in sign of $p'(r)$ or
toroidal effects for $q>1$), $\Delta(Q)$ also has this form, with
$C_{2}\rightarrow-C_{2}$, and the poles corresponding to electrostatic
modes are in the complex plane, and so do not show up on the positive
or negative real $Q$ axis. This calculation does not include the
divergence of the $E\times B$ drift or perpendicular resistivity
$\eta_{\perp}$ (resistive particle flux) included in the early treatments,
and shows that these effects are not important for the qualitative
behavior that is evident from this dispersion relation. This behavior
includes the occurrence of complex roots $Q$ for $\Delta'<\Delta_{min}$
and the stabilization of these roots for $\Delta'<\Delta_{crit}$,
with $0<\Delta_{crit}<\Delta_{min}$. The analysis is aided by the
use of the layer symmetry in the RI regime, namely $p'\rightarrow\nu^{3/2}p',\,\, c_{s}\rightarrow\nu^{3/4}c_{s},\,\,\gamma\rightarrow\nu\gamma$.
In Appendix A we argue that the effects of the divergence of the $E\times B$
drift and the perpendicular resistivity $\eta_{\perp}$ are comparable
in magnitude to the other terms kept; results including these effects
will be included in a future publication. Another point of this analysis
is to illustrate the simple methods that can be employed in the absence
of these last two effects, and to explore further the connection between
the electrostatic resistive interchanges and the non-monotonic behavior
of $\Delta(Q)$ that is responsible for the complex roots.

We have also treated the tearing mode in the VR regime with pressure
gradient and parallel dynamics, again neglecting the divergence of
the $E\times B$ drift and the perpendicular resistivity. The methods
employed are the same as those used in the RI regime. We again find
electrostatic resistive interchanges which cause poles in $\Delta(Q)$,
on the positive real axis for unfavorable curvature and on the negative
real axis for favorable curvature. These calculations are aided by
the symmetry $p'\rightarrow\lambda p',\,\,\gamma\rightarrow\lambda\gamma$
and $c_{s}\rightarrow\lambda c_{s}$, for real (not only positive)
$\lambda$. As the sound speed is increased, the poles of $\Delta'=\Delta(Q)$
on the real axis corresponding to these resistive interchanges move
to $Q=0$ and disappear into the complex plane, except for the last
two. For favorable curvature, these coalesce at negative real $Q$
to form complex conjugate poles with $Q_{r}<0$ rather than disappearing
from the real $Q$ axis into $Q=0$ as in the RI regime. For higher
sound speed, the presence of these last two complex electrostatic
resistive interchange roots is responsible for non-monotonic behavior
of $\Delta(Q)$, with a relative minimum $\Delta_{min}$. For $\Delta'$
below $\Delta_{min}$, the tearing mode dispersion relation $\Delta'=\Delta(Q)$
has complex roots. As $\Delta'$ is lowered further in this range,
these modes with complex frequency are stabilized at $\Delta'=\Delta_{crit}$,
where $0<\Delta_{crit}<\Delta_{min}$. That is, there is a Glasser
effect, with real frequencies as well as stabilization for positive
$\Delta'$, in the VR regime. This non-monotonic behavior exists over
a large range in sound speed.

Tearing modes in two fluid models, including diamagnetic effects,
also typically have real frequencies of order the diamagnetic frequency,
$\omega_{r}\sim\omega_{*}$, with a reduction in growth rate $\gamma$
\cite{Coppi-1,Coppi-2,Biskamp,FMA,ColeFitzpatrick2}. Real frequencies
also occur in tearing modes driven by parallel velocity shear \cite{Finn-vparallel-prime}.
Based on these regimes and the new finite frequency results in the
VR regime in this paper, we can say with confidence that tearing modes
with real frequencies are the rule rather than the exception in most
tearing mode regimes. These frequencies are of course in the plasma
frame; in the laboratory frame these linear modes have an additional
Doppler shift due to $E\times B$ plasma rotation, which is typically
diamagnetic in magnitude.

As we have discussed, these real frequencies are responsible for newly
discovered results related to resonant field amplification of error
fields \cite{FinnColeBrennan} and for a major modification of error
field penetration or locking \cite{FinnColeBrennan}, allowing locking
to just above a finite value of rotation rather than to just above
zero rotation. These effects are related to the behavior of resistive
wall modes for real frequency tearing modes. In Ref.~\cite{FinnGerwinModeCoupling}
it was argued that the \emph{ideal plasma} resistive wall mode is
described as a mode coupling between the plasma mode and the leakage
of flux through the resistive wall. This mode is driven unstable by
$E\times B$ rotation, specifically lowering the critical $\beta$
for instability, because the plasma rotation Doppler shifts the real
frequency stable ideal plasma mode to have small real frequency in
the laboratory frame. A similar, but very weak, destabilization effect
was shown in Ref.~\cite{FinnGerwinModeCoupling} for tearing modes
in the RI regime with no pressure-curvature  drive in the layer. It
was also shown that no such destabilization effect occurs for VR tearing
modes without pressure-curvature drive in the layer, because these
modes are purely growing or purely damped, independent of the sign
of $\Delta'$, in the plasma frame. In this paper we presented a formulation
for the study of resistive wall tearing modes with pressure gradient
and field line curvature in either the RI or VR regime. As we discussed
in an earlier section, the modes with parallel dynamics have real
frequency layers in the plasma frame in both of these regimes. We
have shown a simple model of mode coupling between an ideal wall tearing
mode and an ideal plasma resistive wall mode in the presence of drive
by pressure and curvature in the outer region. We showed that the
most unstable mode is strongly destabilized by slow $E\times B$ rotation
for $\beta$ below the no-wall tearing mode $\beta$ limit, but stabilized
for larger rotation. Also, we found that the destabilizing effect
is maximized when the ideal wall tearing mode has nearly zero frequency
in the laboratory frame, $\omega_{r}+kv\approx0$, maximizing the
coupling of the ideal wall tearing mode and the ideal plasma resistive
wall mode.%
\footnote{A potentially important related result is that of Ref.~\cite{AibaHirota}.
In this paper it was noticed that, similar to the situation in Ref.~\cite{FinnGerwinModeCoupling},
plasma rotation can shift the frequency of a stable ideal MHD mode
such as a TAE mode, to be zero in the laboratory frame. A Doppler
shift of this order in such a mode was found to be destabilizing in
a manner essentially the same as for tearing modes in the present
paper, or for stabilized ideal MHD modes in Ref.~\cite{FinnGerwinModeCoupling}.
This result was surprising because TAE modes are not unstable with
or without a conducting wall.%
}

We have noted a straightforward extension of the resistive wall tearing
results of this paper: We considered double tearing modes with either
or both tearing layers having real frequency layers. Similar considerations
to those pertaining to resistive wall tearing modes suggest a maximum
growth rate at the crossing $\omega_{r1}+kv(r_{1})=\omega_{r2}+kv(r_{2})$.
Such results show that rotation shear, specifically a difference in
plasma rotation between the two mode rational surfaces, can \emph{destabilize}
double tearing modes. Similar considerations apply to poloidal coupling
in toroidal geometry.

\textbf{Acknowledgments.} The work of J.~M.~Finn was supported by
the DOE Office of Science, Fusion Energy Sciences and performed under
the auspices of the NNSA of the U.S.~DOE by LANL, operated by LANS
LLC under Contract No DEAC52-06NA25396. The work of A. J. Cole and
D. P. Brennan was supported by the DOE Office of Science collaborative
Grant Nos. DE-SC0014119 and DE-SC0014005, respectively.

\section*{Appendix A. Divergence of the $E\times B$ velocity and perpendicular
resistivity}

The perturbed perpendicular $E\times B$ flow is $\tilde{\mathbf{v}}_{\perp}=-\nabla\tilde{\Phi}_{es}\times\hat{\mathbf{b}}/B(r)$,
where $\tilde{\Phi}_{es}=-B(r)\tilde{\phi}$ is the perturbed electrostatic
potential. In this extension of reduced MHD, we allow $\hat{\mathbf{b}}$
and $\hat{\mathbf{z}}$ to differ and neither $B_{z}(r)$ nor $B(r)$
is exactly constant. The divergence of $\tilde{\mathbf{v}}_{\perp}$
takes the form 
\[
\nabla\cdot\tilde{\mathbf{v}}_{\perp}\approx2\hat{\mathbf{b}}\times\mathbf{\boldsymbol{\boldsymbol{\kappa}}}\cdot\nabla\tilde{\Phi}_{es}/B\approx-2\nabla\tilde{\phi}\times\hat{\mathbf{z}}\cdot\mathbf{\boldsymbol{\boldsymbol{\kappa}}}.
\]
The equation for the pressure takes the form
\[
\gamma\tilde{p}+\nabla\tilde{\phi}\times\hat{\mathbf{z}}\cdot\left(\nabla p-2\Gamma p\boldsymbol{\mathbf{\boldsymbol{\kappa}}}\right)=0
\]
plus parallel dynamics, so that Eq.~(\ref{eq:AdiabatWithParallel})
of Sec.~\ref{sub:RI-regime-with-parallel} has 
\[
p'(r)\rightarrow p'(r)-2\Gamma p\kappa,
\]
effectively lowering the pressure gradient in Eq.~(\ref{eq:AdiabatWithParallel}).
Note that the pressure gradient term in Eq.~(\ref{eq:vParallelEq})
is unaffected. We see that the pressure gradient term from Eq.~(\ref{eq:AdiabatWithParallel})
and the $\nabla\cdot\tilde{\mathbf{v}}_{\perp}$ term are in the ratio
\[
\frac{r_{t}p'(r_{t})}{\Gamma p}\,\,:\,\,\frac{2B_{\theta}^{2}}{B_{0}^{2}}.
\]
We conclude that the $\nabla\cdot\tilde{\mathbf{v}}_{\perp}$ term,
although it is stabilizing, is small in the cylindrical tokamak limit,
with $B_{\theta}/B\sim r_{t}/R$. (In the RI regime, these two terms
are proportional to the terms in the expression $S-2D_{s}/\Gamma\beta$
of Refs.~\cite{CGJ,FinnManheimer}, in reverse order, and note that
these two terms are comparable if $B_{\theta}\sim B_{z}$, as in an
RFP.) In Sec.~\ref{sub:RI-regime-with-parallel}, in taking $g=G_{0}/b_{0}^{2}$
small, we must not violate this assumption.

Including $\eta_{\perp}\mathbf{j}_{\perp}$ in Ohm's law in the equations
in the tearing layer, we find
\[
\tilde{\mathbf{v}}_{\perp}=-\frac{\nabla\tilde{\Phi}_{es}\times\hat{\mathbf{b}}}{B^{2}}-\eta_{\perp}\frac{\mathbf{\tilde{j}}\times\mathbf{B}}{B^{2}}.
\]
The last term, which approximately equals $-\eta_{\perp}\nabla\tilde{p}/B^{2}$;
$n\tilde{\mathbf{v}}_{\perp}$, is the flux responsible for classical
diffusion (without the Pfirsch Schl\"uter correction). The leading
correction enters the adiabatic law via $\Gamma p\nabla\cdot\tilde{\mathbf{v}}_{\perp}$
and equals $-\eta_{\perp}\Gamma p\nabla^{2}\tilde{p}/B_{0}^{2}=-\eta_{\perp}(\Gamma p/B_{0}^{2})\tilde{p}''$.
To compare its magnitude, we must compare
\begin{equation}
\gamma\,\,:\,\,\frac{\eta_{\perp}\beta}{\delta^{2}}.\label{eq:etaperp::comparison}
\end{equation}
From the RI regime ordering of Sec.~\ref{sec:Resistive-inertial-(RI)-regime},
with $\delta\sim\epsilon$, $\gamma\sim\epsilon^{3/2}$, $\eta_{\perp}\sim\eta_{||}\sim\epsilon^{5/2}$
and $\beta\sim\epsilon$ the terms in Eq.~(\ref{eq:etaperp::comparison})
compare as 
\[
\epsilon^{3/2}\,\,:\,\,\epsilon^{3/2}.
\]
That is, the dominant term proportional to $\eta_{\perp}$ is comparable
to the other terms. Let us consider this same comparison in the VR
regime. Assuming $\delta\sim\epsilon,\,\,\gamma\sim\epsilon^{2},\,\,\eta_{\perp}\sim\eta_{||}\sim\epsilon^{3}$
and $\beta\sim\epsilon$ as in Sec.~\ref{sec:Viscoresistive-(VR)-regime},the
terms in Eq.~(\ref{eq:etaperp::comparison}) scale as 
\[
\epsilon^{2}\,\,:\,\,\epsilon^{2}.
\]
Nevertheless, in either tearing regime it is not inconsistent to ignore
this $\eta_{\perp}$ term, which introduces complexity by increasing
the order of the equations, and the results of Sec.~\ref{sub:RI-regime-with-parallel}
show that the reglect of the $\eta_{\perp}$ term (as well as the
neglect of the $\nabla\cdot\tilde{\mathbf{v}}_{\perp}$ term) do not
make any qualitative difference in the RI regime with pressure-curvature
drive.

\bibliographystyle{plain}

\end{document}